\documentclass[%
 aip,pop, numerical,
 amsmath,amssymb,
 reprint,%
]{revtex4-1}

\usepackage {CJK}

\usepackage{graphicx}
\usepackage{dcolumn}
\usepackage{bm}

\usepackage[utf8]{inputenc}
\usepackage[T1]{fontenc}
\usepackage{mathptmx}
\usepackage{etoolbox}

\makeatletter
\def\@email#1#2{%
 \endgroup
 \patchcmd{\titleblock@produce}
  {\frontmatter@RRAPformat}
  {\frontmatter@RRAPformat{\produce@RRAP{*#1\href{mailto:#2}{#2}}}\frontmatter@RRAPformat}
  {}{}
}%
\makeatother

\begin{document}
\begin{CJK*}{UTF8}{bsmi}

\preprint{AIP/123-QED}

\title[Magnetic Island Merging: Two-dimensional MHD Simulation and Test-Particle Modeling]{Magnetic Island Merging: Two-dimensional MHD Simulation and Test-Particle Modeling}
 
\author{Xiaozhou Zhao (趙小舟)}
 \email{xiaozhou.zhao@kuleuven.be}
 \affiliation{Center for mathematical Plasma Astrophysics, Department of Mathematics, KU Leuven, Celestijnenlaan 200B, 3001 Leuven, Belgium.}%
 
\author{Fabio Bacchini}%
\affiliation{Center for mathematical Plasma Astrophysics, Department of Mathematics, KU Leuven, Celestijnenlaan 200B, 3001 Leuven, Belgium.}%

\author{Rony Keppens}
\affiliation{Center for mathematical Plasma Astrophysics, Department of Mathematics, KU Leuven, Celestijnenlaan 200B, 3001 Leuven, Belgium.}%

\date{\today}

\begin{abstract}
 In an idealized system where four current channels interact in a two-dimensional periodic setting, we follow the detailed evolution of current sheets (CSs) forming in between the channels, as a result of a large-scale merging. A central X-point collapses and a gradually extending CS marks the site of continuous magnetic reconnection. Using grid-adaptive, non-relativistic, resistive magnetohydrodynamic (MHD) simulations, we establish that slow, near-steady Sweet-Parker reconnection transits to a chaotic, multi-plasmoid fragmented state, when the Lundquist number exceeds about $3\times10^4$, well in the range of previous studies on plasmoid instability. The extreme resolution employed in the MHD study shows significant magnetic island substructures. With relativistic test-particle simulations, we explore how charged particles can be accelerated in the vicinity of an O-point, either at embedded tiny-islands within larger "monster"-islands or near the centers of monster-islands. While the planar MHD setting artificially causes strong acceleration in the ignored third direction, it also allows for the full analytic study of all aspects leading to the acceleration and the in-plane-projected trapping of particles in the vicinities of O-points. Our analytic approach uses a decomposition of the particle velocity in slow- and fast-changing components, akin to the Reynolds decomposition in turbulence studies. 
Our analytic description is validated with several representative test-particle simulations. We find that after an initial non-relativistic motion throughout a monster island, particles can experience acceleration in the vicinity of an O-point beyond $\sqrt{2}c/2\approx 0.7c$, at which speed the acceleration is at its highest efficiency
\end{abstract}

\maketitle
\end{CJK*}

\section{Introduction} \label{sec:intro}

Solar eruptions, i.e. coronal mass ejections (CMEs) and flares, are the most energetic phenomena in the solar system.   
It is commonly accepted that the associated solar flares and CMEs are different manifestations of a single physical process.
During a solar flare, up to $10^{32}\,\mathrm{erg}$ of energy is suddenly released~\cite{Benz2008}. It is believed that the flare energy comes from the magnetic energy stored in the solar corona. It is however not well understood how magnetic energy can be released quickly enough to be compatible with observed flaring time scale $\sim100\,\mathrm{s}$. Similar questions are also raised in space and astrophysical research: how is the magnetic energy stored in planetary, stellar, and astrophysical magnetic fields efficiently released?

 The coronal plasma dynamics is described by magnetohydrodynamics (MHD) on macroscopic scales. This follows from the fact that a typical (coronal loop) length scale $L_{0}\simeq 10^{9}\,\mathrm{cm}$ is much larger than the (local) microscopic parameters like the ion Larmor radius ($\sim 10^{2}\,\mathrm{cm}$) and the Debye length ($\sim 2\,\mathrm{cm}$). 
According to resistive MHD theory, the magnetic field in the solar corona with a typical resistivity $\eta\simeq 10^{-16}\,\mathrm{s}$ can be diffused over a time scale of $\tau_{d}\simeq 4\pi L_{0}^{2}/(c^{2}\eta)\simeq10^{14}\,\mathrm{s}$, which is too long to explain observed flare time scale. In contrast, the Alfv\'en transit time scale, i.e., the ratio of the typical length and Alfv\'en speed $t_{\mathrm{A}}\simeq L_{0}/v_{\mathrm{A}}$, is very short in the solar corona, about $1.5\,\mathrm{s}$.
 The relative magnitude of the magnetic diffusion time scale and the Alfv\'en transit time scale is conventionally measured in terms of the Lundquist number, 
 \begin{equation}
 	S_{Lu}=\frac{4\pi L_{0}v_{\mathrm{A}}}{\eta c^{2}}.
 \end{equation} 
 The solar flare time scale $\sim 100\,\mathrm{s}$ is somewhere in-between the Alfv\'en transit time scale $t_{\mathrm{A}}\simeq 1.5\,\mathrm{s}$ and the magnetic diffusion time scale $\tau_{d}\simeq10^{14}\,\mathrm{s}$, which implies that the flare process is neither a pure ideal MHD process nor a pure diffusion process but something mixed.

The Sweet-Parker reconnection model was formulated in Refs.~\onlinecite{sweet1958production,Sweet1958,Parker1957,parker1963solar} where there exists a current sheet (CS) with a length close to the global length scale arising between antiparallel magnetic fields. In this model, plasma diffuses into the CS, along its whole length, at a sub-Alfv\'enic inflow speed. The plasma is accelerated along the sheet, and eventually expelled from its two ends at Alfv\'en speeds. According to the Sweet-Parker model, magnetic energy is released to the plasma via reconnection on a typical time scale of a few tens of days, far too slow to account for flares but much faster than the pure diffusion process. A fast reconnection mechanism called Petschek model was proposed in Ref.~\onlinecite{Petschek1964}, which contains a tiny CS with a width $\Delta\sim O(\eta)$ rather than the system size. The Petschek reconnection configuration is characterized by two pairs of slow-mode shocks far from the neutral point. The two pairs of shocks can be regarded as the extension of the tiny CS~\cite{Priest1982}, where the main conversion of magnetic energy into kinetic and thermal energy takes place. The Petschek reconnection is fast enough to be compatible with the time scales of solar flares. However, it is not self-consistent in MHD regime because an anomalous resistivity is required in the diffusion region. The Sweet-Parker and Petschek reconnection mechanisms belong to the steady MHD models of magnetic reconnection. Except steady reconnection, there are unsteady reconnection mechanisms. 
As shown in Ref.~\onlinecite{dungey1953lxxvi}, a CS can form as a result of the unsteady collapse of an X-point and suggested that the magnetic field lines can be "broken" and become "reconnected". A linear-stability analysis was conducted in Ref.~\onlinecite{Furth1963}, which showed that a sheet pinch is unstable at finite resistivity when the aspect ratio of the sheet is larger than $2\pi$. This resistive tearing instability produces a change in the topology of the magnetic field, creating closed magnetic loops called magnetic islands or "plasmoids". The magnetic null point in the center of a plasmoid is called the O-point. The scenario of fractal reconnection was depicted in Ref.~\onlinecite{ShibataTanuma2001}, where the CS tends to cascade to smaller scales and multiple magnetic islands are formed, presenting fractal structure. During the cascading process, energy is transported from large scales to small scales, and dissipated in small scales.  The demonstration of the fractal structure of the CS needs ultra-resolution simulations and this was done later, e.g., in Refs.~\onlinecite{Barta2011,Ni2012ApJ75820N,Ni2018ApJ}. Besides, a non-linear study of the CS instability and the formation of plasmoid chains was conducted in Ref.~\onlinecite{Loureiro2007}. A fast reconnection model via a stochastic plasmoid chain was proposed in Ref~\onlinecite{Uzdensky2010}, which presents a self-similar distribution of plasmoid sizes and fluxes. The reconnection rate of fractal reconnection is high enough to explain the flare time scale, and is considered as a candidate mechanism for fast reconnection in MHD regime. 
Now the question of the criterion for plasmoid instability arises. Recent resistive MHD simulations demonstrate that a CS can get fragmented into multiple magnetic islands and thinner CSs by the cascading process, and it facilitates the various types of reconnection taking place simultaneously, once the Lundquist number exceeds a critical value~\cite{Bhattacharjee2009, Huang2010PhPl,Shen2011,Mei2012} of about $4\times10^4$.  
The onset of the resulting plasmoid instability leads to a reconnection rate nearly independent of the Lundquist number. Although it is established that the critical Lundquist number to trigger the plasmoid instability is $\sim 10^4$, it is still worthy to check the criterion for fast reconnection in different numerical setups.

Magnetic reconnection is so important because it is ubiquitous in space, astrophysical and laboratory plasmas. We take the standard solar eruption model as an example to illuminate the importance of magnetic reconnection. 
The standard solar eruption model depicts that a CS usually forms and develops between the flare arcade and the CME bubble~\cite{ForbesActon1996,LinForbes2000}, where reconnection takes place. The core of the CME usually corresponds to an eruptive prominence. 
Hard X-ray observations of solar flares show that there are a pair of hard X-ray sources from the footpoints of the flare arcade in the chromosphere as well as an above-the-loop-top hard X-ray source~\cite{Masuda1994}. The hard X-ray radiation from flares is usually interpreted as the bremsstrahlung radiation from the energetic particles accelerated during the flare process. It is however not clear where and how these particles are accelerated during reconnection processes. Particle acceleration during reconnection provides important clues to understand the acceleration and transport of energetic particles in the Universe. In this paper, we investigate the criterion for plasmoid instability and the role of plasmoids during particle acceleration. 
The complexity for the particle acceleration process in reconnection is twofold: firstly, the reconnection itself is complex; secondly, the electromagnetic field associated with reconnection, in which the particles are accelerated, is complex. 
Several works have investigated these mechanisms with test-particle simulations in MHD magnetic reconnection.
The particle acceleration by the induced electric fields and magnetic gradient and curvature drift effects in cascading reconnection in the framework of the guiding-center approximation (GCA, also called adiabatic approximation) was studied in Refs.~\onlinecite{Zhou2015xwApJ1,Zhouxw2015ApJ2}. The particle motion in a system with two parallel, repelling current channels as a simplified representation of flux ropes was studied with relativistic test particle methods in Refs.~\onlinecite{Ripperda2017MNR3279R,Ripperda2017MNR3465R}.
The mechanism for accelerating electrons in contracting magnetic islands was studied in Refs.~\onlinecite{Drake2006,Guidoni2016,Roux2019ApJ}.
Such studies about the particle acceleration process in plasmoids usually treat the particle motion in the framework of adiabatic motion. 
The non-adiabatic motion of particles inside the plasmoids requires further investigations, which is however a fundamental step to take for the following reasons: (1) the conditions for adiabatic motions are not always preserved, and non-adiabatic motions do exist in plasmas; (2) particles may be energized by non-adiabatic processes, which should be confirmed by numerical and analytical studies. 
In order to do so, the test particle orbits should be fully resolved with a very high-resolution MHD background, which, in this paper, we are going to investigate.

The paper is organized as follows. Section~\ref{sec:model} briefly describes the resistive MHD model. Section~\ref{sectestparticle} shows the results of test-particle simulations with an emphasis on a selected representative particle. Section~\ref{otherparticle} briefly describes the motion of particles other than the representative particle. Section~\ref{analytical} develops an analytical method to interpret the numerical results of particle acceleration.
Section~\ref{discussion} briefly discusses the limitations and implications of this study. Section~\ref{conclusion} summarizes the paper.

\section{MHD simulations} \label{sec:model} 
\subsection{The governing equations}

\begin{table*}[ht!]
	\begin{center}
\caption{Normalization Units.\label{Units}}
        {
\begin{tabular}{r r r r}
\hline\hline
Symbol & Quantity & Unit & Value\\
$x,y,z$ &Length & $L_{0}$ & $10^{9}\,\mathrm{cm}$\\

$T$ &Temperature & $T_{0}$ &  $1.0\times 10^{6} \,\mathrm{K}$ \\
$n$ & Number density & $n_{0}$ & $1.0\times 10^{9} \,\mathrm{ cm^{-3}}  $\\
$\rho$ & Mass density & $\rho_{0}=1.4n_{0}m_{\mathrm{H}}$ & $2.3417\times 10^{-15} \,\mathrm{g\cdot cm^{-3}}  $\\
$p$ &Pressure & $p_{0}=(\rho_{0} k_{B} T_{0})/(\mu_{w} m_{\mathrm{H}}) $ & $0.3175 \, \mathrm{Ba}$ \\
$\mathcal{H}$ & Energy density&  $p_{0}$ & $0.3175 \, \mathrm{erg\cdot cm^{-3}}$ \\
$\mathbf{B}$ &Magnetic induction& $B_{0}=\sqrt{4\pi p_{0}}$ & $1.9976\,\mathrm{ Gauss}$ \\
$\mathbf{u},\mathbf{v}$ &Velocity & $v_{0}=B_{0}/\sqrt{4\pi \rho_{0}} $ & $1.1645 \times 10^{7} \,\mathrm{cm\cdot s^{-1}}$\\

$t$ &Time&  $t_{0}=L_{0}/v_{0}$ &$85.8746 \,\mathrm{s}$ \\

$\eta$ & Resistivity&  $\eta_{0}=(4\pi L_{0}^{2})/(c^{2}t_{0})$ & $1.6282\times 10^{-4} \,\mathrm{s}$ \\

$\mathbf{E}$ & Electric field&  $(B_{0}L_{0})/(t_{0}c) $ & $7.759\times 10^{-4}\, \mathrm{statvolt/cm}$ \\

$\mathbf{J}$ & Current density&  $(B_{0}c)/(4\pi L_{0}) $ & $4.7689\,\mathrm{statamp/cm^{2}}$ \\
\hline
\end{tabular}}
	\end{center}
\end{table*}  
 
We solve the following set of resistive MHD equations:
\begin{equation}\label{eq:1}
\frac{\partial\rho}{\partial t}+\nabla\cdot \left(\rho\mathbf{u}\right)=0,
\end{equation}

\begin{equation}\label{eq:2}
\frac{\partial\left(\rho\mathbf{u}\right)}{\partial t}+\nabla\cdot \left[\rho\mathbf{u}\mathbf{u}+\left(p+\frac{\mathbf{B}^{2}}{8\pi}\right)\mathbf{I}-\frac{\mathbf{B}\mathbf{B}}{4\pi}\right]=  \mathbf{0},
\end{equation}

\begin{equation} \label{eq:3}
\frac{\partial \mathcal{H}}{\partial t}+\nabla\cdot \left[\left(\mathcal{H}-\frac{\mathbf{B}^{2}}{8\pi}+p\right)\mathbf{u}+\frac{c\mathbf{E}\times\mathbf{B}}{4\pi}\right]= 0,
\end{equation}

\begin{equation}\label{eq:4}
\frac{\partial\mathbf{B}}{\partial t}+\nabla\cdot \left(\mathbf{u}\mathbf{B}-\mathbf{B}\mathbf{u}  \right)= -c\nabla\times(\eta\mathbf{J}),
\end{equation}
where $\mathcal{H}=[\rho\varepsilon+(1/2)\rho\mathbf{u}^{2}+\mathbf{B}^{2}/8\pi]$ is the total energy density, $\varepsilon=p/(\gamma\rho-\rho)$ is the internal energy per unit mass, $\gamma$ is the adiabatic index, $\mathbf{E}=[\eta\mathbf{J}-(\mathbf{u}\times\mathbf{B})/c]$ is the electric field, $\mathbf{J}=(c/4\pi)\nabla\times\mathbf{B}$ is the electric current density, $\eta$ is the resistivity, and $c$ is the light speed in vacuum. In this paper, $\mathbf{u}$ represents the fluid bulk velocity while we will further on use the symbol $\mathbf{v}$ for the test particle velocity.
In numerical simulation, the above equations are solved in dimensionless form.  
To non-dimensionalize the equations, each variable is divided by its normalizing unit. 
The normalizing units of variables are given in Table~\ref{Units}, and the units of other derived variables are listed as well. In this paper, we retain all physical units.
The CGS-Gaussian units are used throughout the paper.
The set of MHD equations is closed by the equation of state
\begin{equation}
	p=\frac{\rho k_{\mathrm{B}}T}{\mu_{w} m_{\mathrm{H}}},
\end{equation}
where $m_{\mathrm{H}}$ is the hydrogen atom mass and the mean molecular weight $\mu_{w} = 1.4/2.3$ for the fully ionised plasma with a $10:1$ abundance of hydrogen and helium.

\subsection{The initial and boundary conditions}
The following two-dimensional magnetic field (e.g., Chapter 2, Page 86 in Ref.~\onlinecite{PriestForbes2000}) is adopted as the initial magnetic topology:
\begin{equation}
{B}_{x}=A\sin\left(\frac{2\pi x}{L_{0}} \right)\cos\left(\frac{2\pi y}{L_{0}} \right), 
\end{equation} 
\begin{equation}
{B}_{y}=-A\cos\left(\frac{2\pi x}{L_{0}} \right)\sin\left(\frac{2\pi y}{L_{0}} \right),
\end{equation} 
where $A=10 B_{0}$. The quantities $B_{0}=1.9976\,\mathrm{Gauss}$ and $L_{0}= 10^{9}\,\mathrm{cm}$ are the measures of length and magnetic field strength respectively, as listed in Table~\ref{Units}. The simulation domain is a square box covering the region $[-0.5L_{0},0.5L_{0}]\times [-0.5L_{0},0.5L_{0}]$ in the {\it x-y} plane.

The initial pressure is obtained by solving the force equilibrium condition
 \begin{equation}
 	-\nabla p+\frac{1}{c} \mathbf{J}\times \mathbf{B}=\mathbf{0}.
 \end{equation}
The analytical expression for the initial pressure is 
\begin{equation}
\begin{split}
p&=\frac{\mathcal{P}}{4\pi} -\frac{A^{2}}{16\pi}\cos\left(\frac{4\pi y}{L_{0}}\right)\\&
-\frac{A^{2}}{8\pi} \left[\sin\left(\frac{2\pi y}{L_{0}}\right)\right]^{2}\cos\left(\frac{4\pi x}{L_{0}}\right),	
\end{split}
\end{equation}
 where $\mathcal{P}=40p_{0}$.
The initial temperature is taken as $T=8T_{0}$ everywhere. Here $p_{0}$ and $T_{0}$ are listed in Table~\ref{Units}.   The plasma-$\beta$ given by our setup is $\beta\simeq 0.8$, close to but less than unity. This plasma-$\beta$ value corresponds to the high corona at a height about $10L_{0}=100\,\mathrm{Mm}$ above the photosphere~\cite{Gary2001SoPh}. The initial mass density is obtained by using the equation of state.

The initial values of the two components of the momentum density $\rho \mathbf{u}$ are set as follows, 
\begin{equation}
	\rho u_{x}=M \sin\left( \frac{2\pi y}{L_{0}}\right) ,
\end{equation} 
and
\begin{equation}
	\rho u_{y}=M \sin\left( \frac{2\pi x}{L_{0}}\right) ,
\end{equation}     
where $M=0.05\rho_{0}v_{0}$ has the dimension of momentum density, and $\rho_{0}$ and $v_{0}$ are listed in Table~\ref{Units}.

\begin{figure}[ht!]
\includegraphics[scale=0.3]{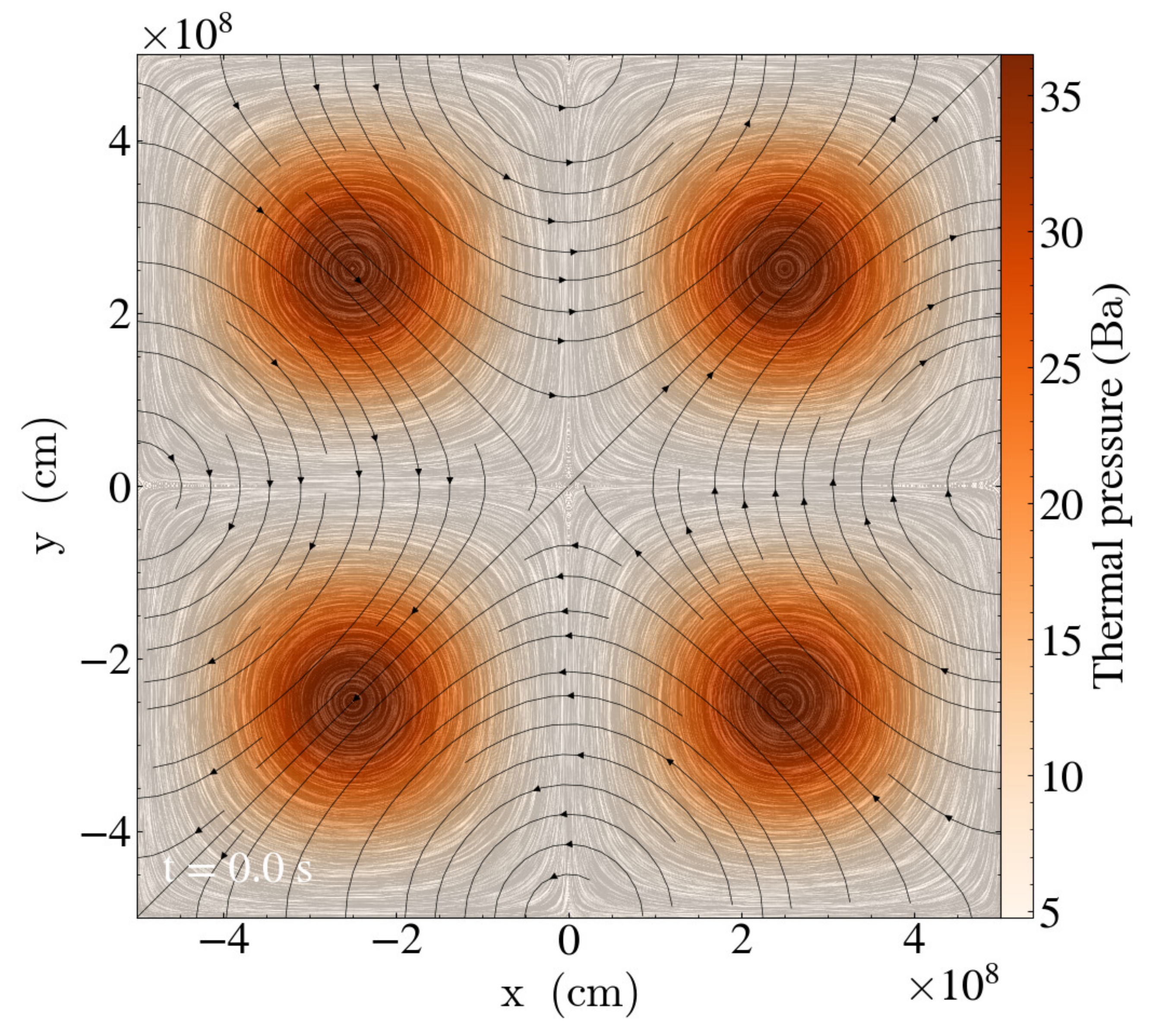}
\caption{\label{initial} The initial thermal pressure distribution with the initial magnetic field overlaid (in grey). The initial velocity field is overlaid as black lines with arrows.    }
\end{figure}

The initial thermal pressure distribution is plotted in Fig.~\ref{initial} with the initial magnetic field overlaid by the line integral convolution (LIC) technique~\cite{cabral1993imaging} integrated in {\it yt}~\cite{YT2011ApJS}, a toolkit for analyzing and visualizing quantitative data. The initial velocity field is overlaid as black arrows. Four magnetic islands are initially present in the simulation domain, and an X-point is located amidst them in the center of the domain.

\subsection{MHD runs}
 
\begin{table*} [ht!]
	\begin{center}
\caption{Numerical experiments.\label{tests}}
        {
\begin{tabular}{r| r| r| r| r| r }
\hline\hline
Case & Resistivity & AMR levels & Effective resolution & Beginning time for resistive instabilities & Critical Lundquist number \\
Case 6A & $1\times10^{-6}\eta_{0}$ & 9 & $16384\times16384$& $t=9.4\,\mathrm{s}$ & $8.6\times 10^{5}$\\

Case 6B & $1\times10^{-6}\eta_{0}$ & 11& $65536\times65536$& $t=9.4\,\mathrm{s}$ &  $8.4\times 10^{5}$ \\
Case 6C & $1\times10^{-6}\eta_{0}$ & 13 & $262144\times262144$& $t=9.4\,\mathrm{s}$& $8.5\times 10^{5}$\\
Case 5 & $1\times10^{-5}\eta_{0}$ & 9 & $16384\times16384$& $t=10.3\,\mathrm{s}$& $9.5\times 10^{4}$\\
Case 5.5 & $5\times10^{-5}\eta_{0}$ & 11 & $65536\times65536$& $t=15.0\,\mathrm{s}$& $2.9\times 10^{4}$\\
Case 4 & $1\times10^{-4}\eta_{0}$ & 9 & $16384\times16384$& None & None\\
Case 3 & $1\times10^{-3}\eta_{0}$ & 9 & $16384\times16384$& None & None\\
\hline
\end{tabular}}
	\end{center}
\end{table*}

We conduct seven numerical experiments with various resistivities and resolutions. The seven cases are listed in Table~\ref{tests}. The base resolution of all cases consists of $64\times 64$ cells. The higher resolutions are achieved by the parallel, adaptive mesh refinement (AMR) technique~\cite{KEPPENS2012718} for MHD incorporated in {\it MPI-AMRVAC}~\cite{Keppens2012,Porth2014ApJS,Xia2018ApJS,keppens2021mpi}. The block refinement criterion is based on momentum density using a L\"ohner type estimator~\cite{Lohner1987}, which evaluates the variations of a specific variable by calculating discrete second derivatives.
The block refinement ratio is fixed as 2, i.e., a block is split into 4 children blocks in this two-dimensional simulation once the refinement is triggered. By using the AMR technique, small-scale structures can be resolved, e.g., the smallest cell in Case 6A corresponds to a physical size of about $6.1\times10^{-5}L_{0}=610\,\mathrm{m}$, and it is $3.8\times10^{-6}L_{0}=38\,\mathrm{m}$ in Case 6C.
In all cases we apply a finite-volume scheme setup combining the HLL solver with cada-type limiter~\citep{Cada2009JCoPh} for reconstruction, and a three-step Runge-Kutta time integration. 

Due to our choice of initial conditions (see Fig.~\ref{initial}), the two islands located in the upper-left and bottom-right are pushed towards each other and merge by the initial velocity field while the other two islands are pushed away from one another. The initial velocity field triggers the collapse of the X-point and leads to the formation of a CS as depicted in Ref.~\onlinecite{dungey1953lxxvi}. The CS grows in length as the merging of the islands continues. This CS is unstable to the plasmoid instability, which occurs once the Lundquist number exceeds a critical value. The critical Lundquist number given in Ref.~\onlinecite{Huang2010PhPl} is $4\times10^{4}$. This critical condition is obtained in a specific numerical setup, which is still valuable to be checked in different numerical codes and setups. Thus the critical condition for plasmoid instability is reinvestigated in our paper.
The critical Lundquist numbers as well as the beginning time for plasmoid instability for all seven cases are listed in Table~\ref{tests}. According to Table~\ref{tests}, the minimum Lundquist number required to trigger plasmoid instability is $2.9\times10^{4}$, which is consistent with the results in Ref.~\onlinecite{Huang2010PhPl}. However, we also note that higher Lundquist number is required for smaller resistivities. The smaller the resistivity is, the earlier the plasmoid instability starts.

 \subsection{Convergence tests}
\label{Convergence_tests}

\begin{figure*}[ht!]
\includegraphics[scale=0.14]{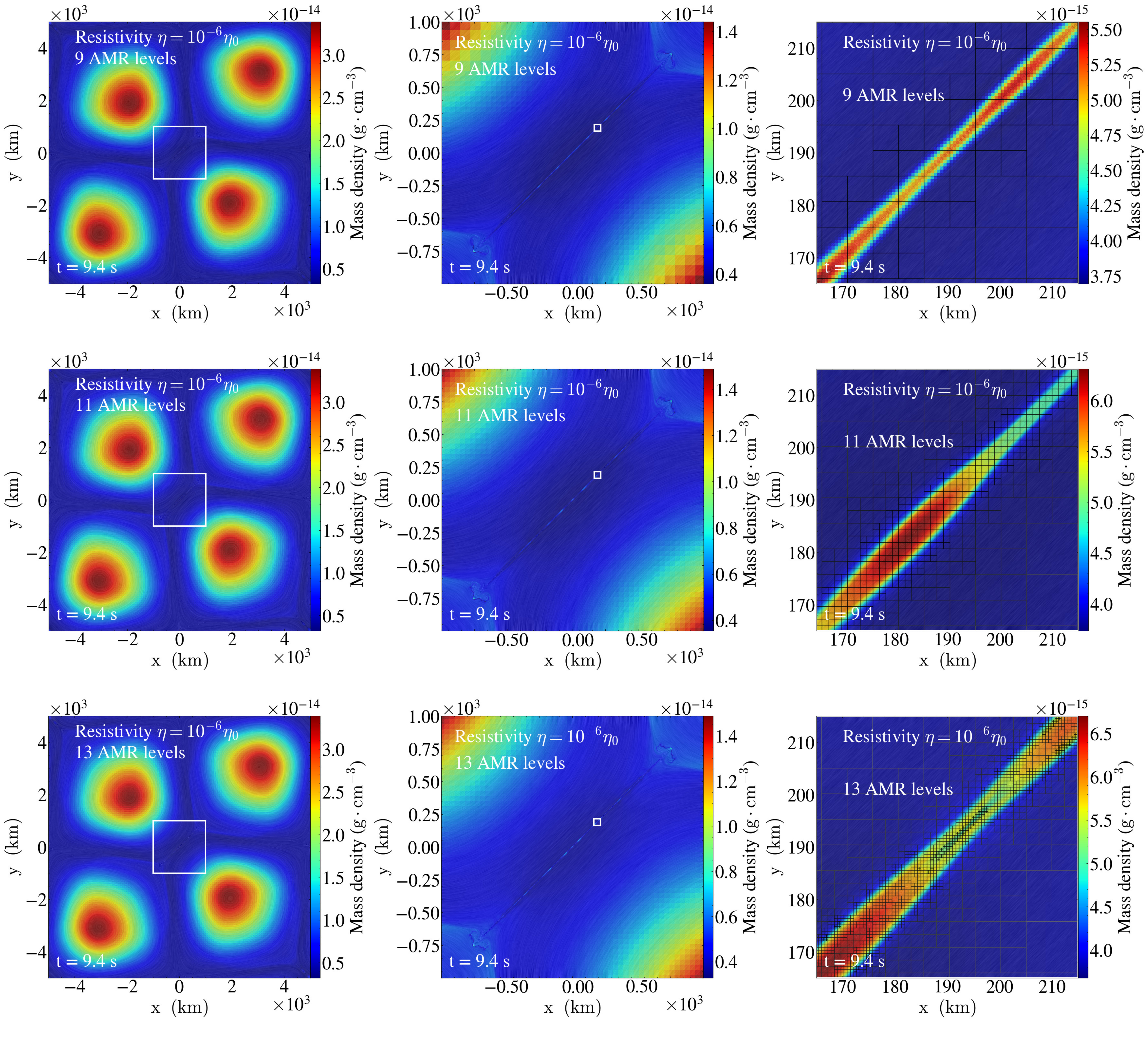}
	\caption{The mass density distributions of Cases 6A (upper row), 6B (middle row) and 6C (bottom row) with magnetic field overlaid. The left column shows the global density distributions, the middle column is the close-up views of the region $[-0.1L_{0},0.1L_{0}]\times[-0.1L_{0},0.1L_{0}]$ indicated by the white boxes in the left column, and the right column is the close-up views of the region $[0.0165L_{0},0.0215L_{0}]\times[0.0165L_{0},0.0215L_{0}]$ indicated by the white boxes in the middle column, respectively. The tree structures of the block-AMR are plotted in the right column as black grids. \label{convergence}}
\end{figure*} 

 A sufficiently high resolution in the simulation is required to reach convergence in results, i.e., the results remain unchanged with higher resolutions. 
To confirm that our resolution is sufficient to resolve the CS, three runs, Cases 6A, 6B and 6C, are carried out with the same resistivity $\eta=10^{-6}\eta_{0}$ but different resolutions. The mass density distribution at $9.4\,\mathrm{s}$ for the three cases are plotted in Fig.~\ref{convergence} with the magnetic field overlaid. The top panels show the results for Case 6A with 9 AMR levels, the middle panels show the results for Case 6B with 11 AMR levels, and the bottom panels show the results for Case 6C with 13 AMR levels. The first column shows the whole simulation domain. The second column shows close-up views of the region $[-0.1L_{0},0.1L_{0}]\times[-0.1L_{0},0.1L_{0}]$, which is indicated by the white boxes in the first column. The third column shows close-up views of the region $[0.0165L_{0},0.0215L_{0}]\times[0.0165L_{0},0.0215L_{0}]$ indicated by the white boxes in the second column. The tree structures of the block-AMR in the simulation are plotted in the third column as black boxes. A block in our simulations consists of $8\times8$ cells. The tree structure of the block-AMR of Case A clearly shows that the thickness of the CS is resolved by about 8 cells, i.e. the size of one grid block as shown. The CSs in Cases B and C are resolved by about 32 cells and 160 cells respectively.

 Although the precise locations and the thicknesses of the islands in Cases B and C are slightly different from Case A, there is no further fragmentation of the CS below the scale already resolved at 9 AMR levels, i.e., no extra islands exist between islands already resolved at 9 AMR levels versus 11 and 13 levels.  
The structures of the CSs in all three cases are qualitatively the same in the sense of the non-existence of extra islands, the same global configurations and the same start time for the plasmoid instability. Thus we confirm that the CS is sufficiently resolved at 9 AMR levels, and further refinement does not alter the structure of the CS qualitatively. 
 
 \subsection{Reconnection type versus resistivity }

\begin{figure*}
\includegraphics[scale=0.25]{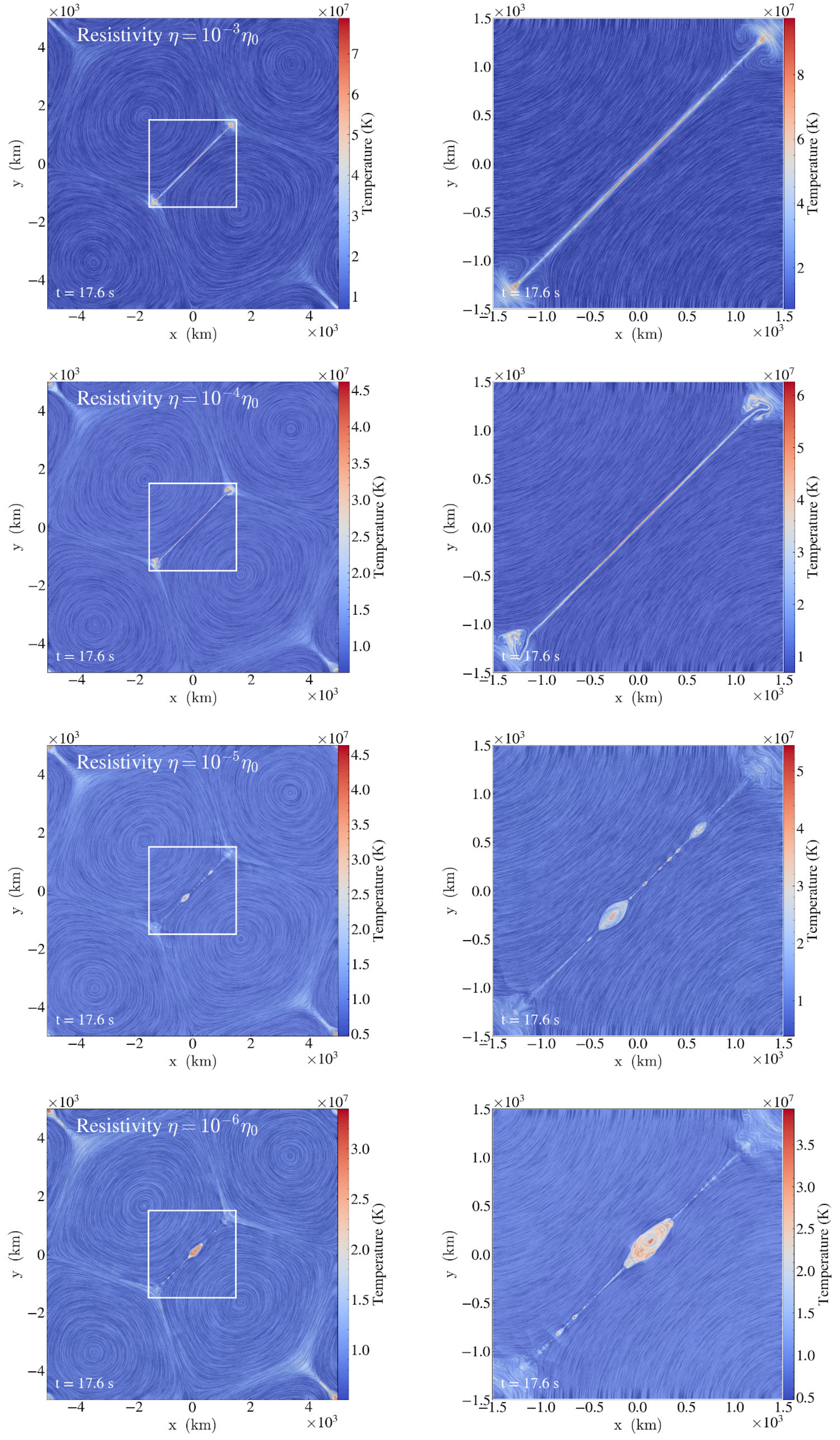}
	\caption{From top to bottom, the temperature distributions of Cases 3, 4, 5, and 6A are plotted,  respectively, with the magnetic field lines overlaid. The right panels are close-up views of the region $[-0.15L_{0},0.15L_{0}]\times [-0.15L_{0},0.15L_{0}]$ that is inside the white boxes in the left column.   \label{resistiveariation}}
\end{figure*}

The temperature distributions for Cases 3, 4, 5, and 6A at $t=17.6\,\mathrm{s}$ are plotted in Fig.~\ref{resistiveariation} with magnetic field lines overlaid. The resolution of the 4 cases are the same while the resistivities are different as listed in Table~\ref{tests}. The left column shows the temperature distributions in the whole domain and the right column shows close-up views of the regions inside the white boxes in the left column. The plasmoid instability occurs in Cases 5 and 6A but does not occur in Cases 3 and 4. The CS in Cases 3 is symmetric while symmetry breaking arises as the resistivity decreases in Case 4. The reconnection in Cases 3 and 4 is the typical Sweet-Parker reconnection. 
Multiple magnetic islands appear in Cases 5, 5.5 and 6A after the plasmoid instability starts. The results of Case 5.5 are not shown here because they are similar to Cases 5 and 6A but with fewer magnetic islands.
It should be noted that the temperature in most of the simulation domain is close to the typical coronal temperature ($\sim 1\,\mathrm{MK}$) while the temperature in several small regions inside the islands is an order of magnitude higher due to resistive heating.

   \subsection{Reconnecting electric field} 
\begin{figure*}[ht!]
 
    \includegraphics[scale=0.2]{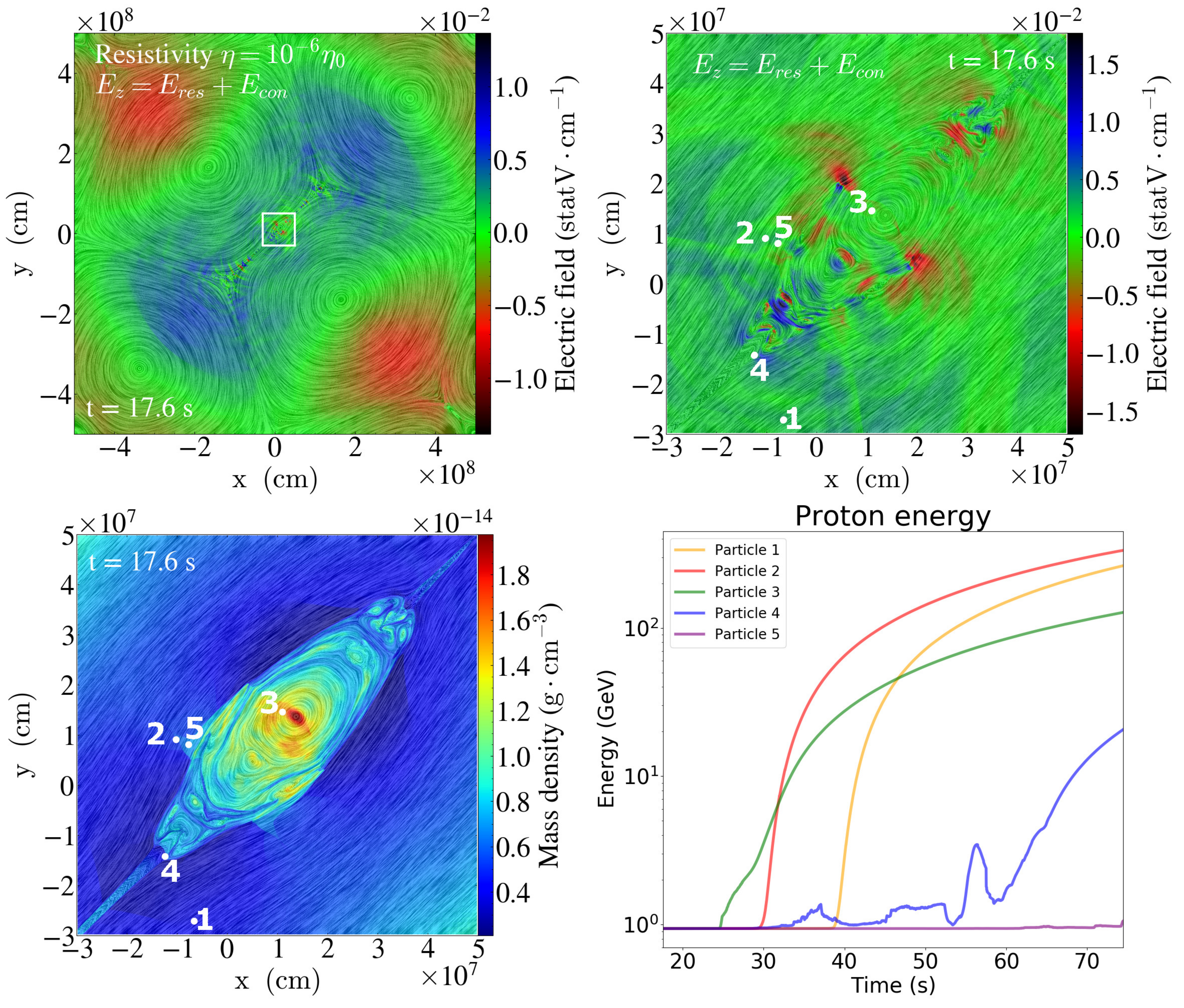}
	\caption{The total electric field $E_{z}=E_{res}+E_{con}$ distribution in the whole simulation domain is plotted in the top-left panel. The top-right panel shows $E_{z}$ in the region $[-0.03L_{0},0.05L_{0}]\times [-0.03L_{0},0.05L_{0}]$ represented by the white box in the top-left panel. The bottom-left panel shows the mass density in the same region. The magnetic field lines are overlaid on each panel. The initial positions of the 5 test particles are plotted as white dots in the top-right and bottom-left panels. The bottom-right panel shows the relativistic energy of the 5 particles versus time in log scale.\label{partinit} }
\end{figure*}

 The electric field calculated from the MHD simulation is 
 \begin{equation}
 	\mathbf{E}=\eta \mathbf{J}-\frac{1}{c}\mathbf{u}\times \mathbf{B}.
 \end{equation}
Due to our two-dimensional setup, $\mathbf{u}$ and $\mathbf{B}$ have only the {\it x} and {\it y}-components. Thus $\mathbf{u}\times \mathbf{B}$ and $\mathbf{J}=(c/4\pi)\nabla\times \mathbf{B}$ are directed along the {\it z}-axis, and so is $\mathbf{E}$. This {\it z}-component of $\mathbf{E}$ reads
\begin{equation}
	E_{z}=\eta J_{z}+\frac{1}{c}(u_{y}B_{x}-u_{x}B_{y}),
\end{equation}
where $\eta J_{z}$ is the resistive term and $(1/c)(u_{y}B_{x}-u_{x}B_{y})$ is the convective term. The upper-left panel of Fig.~\ref{partinit} shows the distribution of $E_{z}$ in the region $[-0.5L_{0},0.5L_{0}]\times[-0.5L_{0},0.5L_{0}]$ with magnetic field lines overlaid. The upper-right panel is a close-up view of the region $[-0.03L_{0},0.05L_{0}]\times[-0.03L_{0},0.05L_{0}]$, which is indicated by the white box in the upper-left panel. 
The bottom-left panel shows the mass density distribution in the region $[-0.03L_{0},0.05L_{0}]\times[-0.03L_{0},0.05L_{0}]$ with magnetic field lines overlaid. A large magnetic island, termed as the "monster" island, is present in the region $[-0.03L_{0},0.05L_{0}]\times[-0.03L_{0},0.05L_{0}]$ with complex structure. As shown in the bottom-left panel, the two ends of the monster island interact with the reconnecting outflows, presenting chaotic patterns. Multiple small islands can be found inside the monster island. 
 The ratio of the convective term $E_{con}=(1/c)(u_{y}B_{x}-u_{x}B_{y})$ and the resistive term and $E_{res}=\eta J_{z}$ is equal to the magnetic Reynolds number $R_{m}$, i.e.,
\begin{equation}
\frac{|E_{con}|}{|E_{res}|}=\frac{|(\mathbf{u}\times\mathbf{B})/c|}{|[(\eta c)/(4\pi)]\nabla\times \mathbf{B}|}=\frac{4\pi}{c^{2}}\frac{ul}{\eta}=R_{m},   \label{Reynolds0}
\end{equation}
where $u$ is the typical fluid speed and $l$ is the typical length scale. Here we point out that the magnetic Reynolds number $R_{m}$ calculated based on the fluid speed $u$ is different from the Lundquist number $S_{Lu}$ calculated based on the Alfv\'en speed. The global magnetic Reynolds number $R^{G}_{m}$ of the system is estimated by taking $l$ and $u$ as the typical global length scale $l=L_{0}$ and the characteristic reconnection inflow speed $u=0.1v_{0}$, respectively, where $L_{0}$ and $v_{0}$ are listed in Table~\ref{Units}. The global magnetic Reynolds number is much larger than unity, i.e.,
\begin{equation}
R^{G}_{m}=\frac{|E_{con}|}{|E_{res}|}\simeq 10^{5}\gg1.   \label{Reynolds}
\end{equation}
 So the total electric field $E_{z}$ is dominated by the convective term $E_{con}$ in most of the simulation domain. The fundamental process of MHD reconnection occurs in an almost-deal but resistive plasma whose global magnetic Reynolds number is much larger than unity. 


 \section{Test-particle simulations}  
    \label{sectestparticle}
\begin{table*} [ht!]
	\begin{center}
\caption{Phases for the motion of proton 2.\label{particlephase}}
        {
\begin{tabular}{r| r| r| r| r| r }
\hline\hline
Phases & Duration & Gyroradius & Speed & Newton vs. Einstein & Features \\
Phase 1 & $17.6-18.6\,\mathrm{s}$ & $r_{g}\sim 10^{-4}\,\mathrm{km}$ & $v\sim 10\,\mathrm{km\cdot s^{-1}}$ & Newtonian dynamics & Guiding center drift\\
Phase 2 & $18.6-27.7\,\mathrm{s}$ & $r_{g}\sim 10^{-2}\,\mathrm{km}$ & $v\sim 0.01c$ & Newtonian dynamics & Adiabatic motion\\
Phase 3 & $27.7-18.6\,\mathrm{s}$ & $r_{g}\sim 10^{2}\,\mathrm{km}$ & $v\sim c$ & Relativistic dynamics & Non-adiabatic motion\\
\hline
\end{tabular}}
	\end{center}
\end{table*}

Among all the MHD simulation cases listed in Table~\ref{tests}, we use Case 6A to do test particle simulations because the plasmoid instability occurs and the numerical resolution of this case, which is the lowest among all cases though, is high enough to resolve the CS as discussed in Section~\ref{Convergence_tests}. This MHD simulation only provides the background electromagnetic field, and there is no feedback from particles to the background field.
A charged particle of charge $q$ and mass $m$ in a given electromagnetic field evolves in time according to the relativistic equation of motion
\begin{equation}
	\frac{\mathrm{d}\mathbf{p}}{\mathrm{d}t}=q\left(\mathbf{E}+\frac{1}{c}\mathbf{v}\times\mathbf{B}\right) \label{motion}
\end{equation}
where $\mathbf{p}=\gamma m\mathbf{v}$ is the relativistic momentum and $\gamma=1/\sqrt{1-(v/c)^{2}}$ is the Lorentz factor. The velocity $\mathbf{v}$ is the time derivative of the position of the particle 
\begin{equation}
	\mathbf{v}=\frac{\mathrm{d}\mathbf{r}}{\mathrm{d}t}.\label{vdefi}
\end{equation}
The solution of the above equations for the Lorentz dynamics is carried out numerically by a fully-implicit iterative scheme which was shown earlier to introduce no numerical errors in the particle energy and can, for this reason, be employed in energy-conserving full-particle simulations \cite{LapentaMarkidis2011APS,Ripperda2018ApJS,Bacchini2019JPhCS1225a2011B}.  The electromagnetic fields at particle positions are obtained by linear interpolation from the MHD simulation grids. 

We run all our test-particle simulations in a fixed fluid snapshot, i.e., the dynamics of the particles is governed by the relativistic equations of motion while the MHD background is kept fixed in time. The fluid snapshot at $17.6\,\mathrm{s}$, which was shown in the bottom panels of Fig.~\ref{resistiveariation}, is taken as the fixed MHD background. The initial time for the test particle runs is arbitrarily set to equate the fluid snapshot, i.e. $17.6\,\mathrm{s}$.  Five protons are placed into the system with zero initial velocities. The initial energies of the five protons are thus the rest energy of protons $938\,\mathrm{MeV}$. The initial positions of the five protons are marked by the white dots in the upper-right panel and the bottom-left panel of Fig.~\ref{partinit}, where each particle is associated with a number.

\subsection{Energetics} 
The time variations of the relativistic energy $\mathcal{E}=\gamma mc^{2}$ of the 5 protons are plotted in the bottom-right panel of Figure~\ref{partinit} in log scale. The relativistic energies of protons 1-4 reach tens to hundreds of GeVs in tens of seconds. The energy of proton 5 does not change too much because the simulation stops exactly at the very beginning of its acceleration. In order to understand the process of proton acceleration, we study the motion of proton 2 in detail.

 \subsubsection{Conditions for adiabaticity}\label{Adiabatic}
 The so-called adiabatic invariants~\cite{Northrop1966} remain almost constant during the motion when the (spatial or temporal) variation of the electromagnetic fields is slow with respect to the typical (space or time) scales of the particle gyration.
  The electromagnetic field in our simulation is fixed in time, hence the variation of the field only comes from spatial gradients as the particle moves through the computational grid. Therefore, the instantaneous field acting on the particle is considered as a function of time via the position of the particle although the field itself is stationary in our simulation.  
 The condition for the slow variation of a field $\lambda$ means that $\lambda$ only changes slightly during the period of the gyration $T_{g}$, i.e.,
\begin{equation}
	\dot{\lambda}T_{g}\ll \lambda. \label{adiabaticond}
\end{equation}
 The so-called third adiabatic invariant, i.e., the total magnetic flux $\Phi$ enclosed by the orbits of gyration of radius $r_{g}$, exists if the condition for adiabaticity indicated by inequality~(\ref{adiabaticond}) is satisfied.
 If the field $\lambda$ were strictly homogeneous, the adiabatic invariants would be strictly conserved. When the field varies only lowly but satisfying inequality~(\ref{adiabaticond}), the adiabatic invariants are not strictly conserved but the rate of change of the adiabatic invariants will also be small \citep{Landau1975} and can be regarded as "quasi-conserved".

\subsubsection{Motion of Particle 2}\label{particle_motion}

 \begin{figure*} [p]
	
    \includegraphics[scale=0.45]{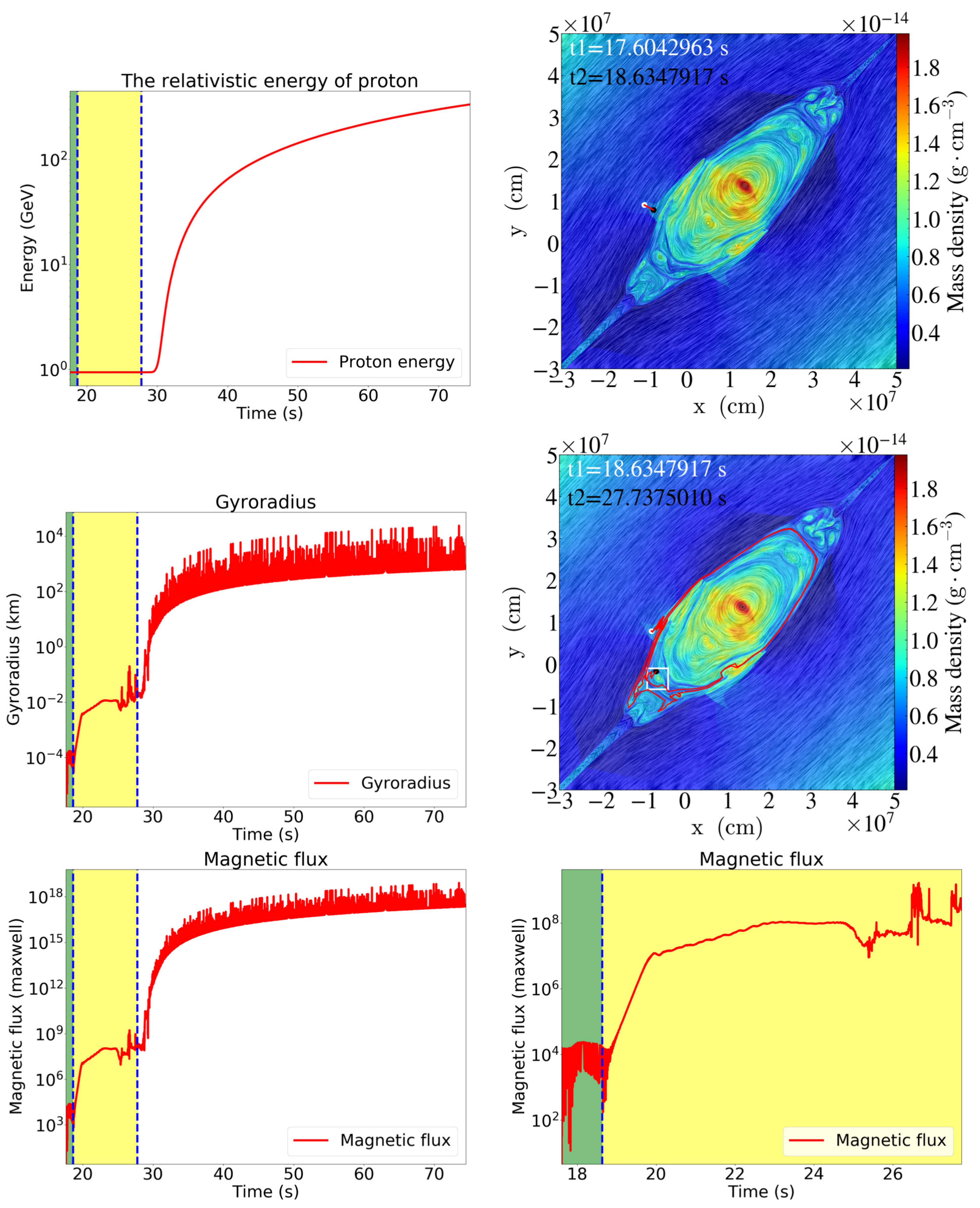}
	\caption{From top to bottom, the relativistic energy, the gyroradius and the magnetic flux enclosed by the orbits of Particle 2 are plotted versus time in log scale in the left column. The periods of Phase 1 and Phase 2 are shaded by green and yellow, respectively. The projected trajectory of Particle 2 in Phase 1 and 2 are plotted by red curves in the upper and middle-right panels with the start and end points marked by white and black dots respectively. A tiny island exists inside the white box in the middle-right panel. The bottom-right panel is a close-up view of the curve of the magnetic flux through the orbit of Particle 2 in Phase 1 and 2.     \label{single}}
\end{figure*}

 \begin{figure*} [p]
	
    \includegraphics[scale=0.5]{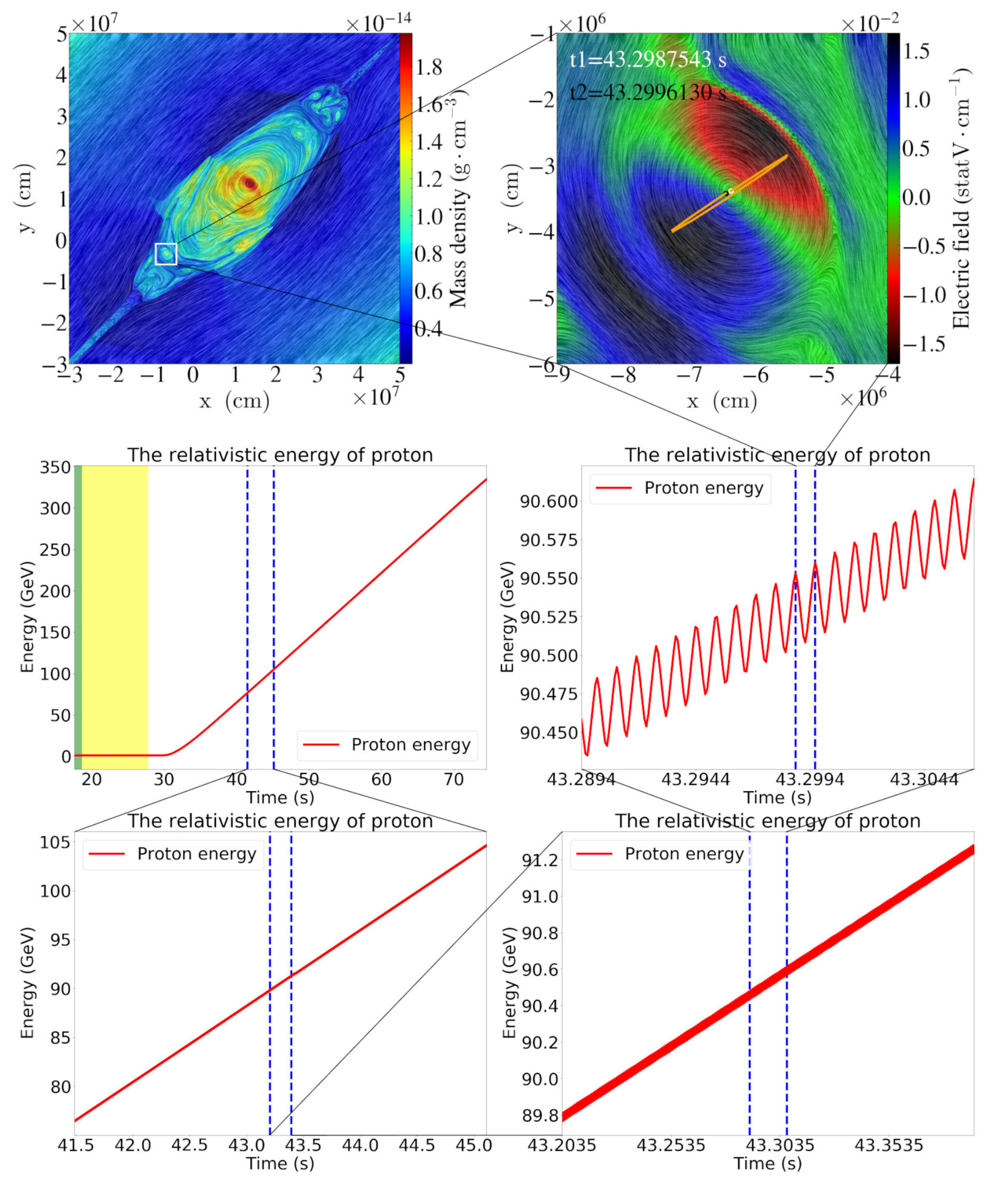}
	\caption{The top-left panel shows the mass density distribution in the region $[-0.03L_{0},0.05L_{0}]\times [-0.03L_{0},0.05L_{0}]$ overlaid with magnetic field lines. The top-right panel shows the distribution of electric field inside the white square shown in the top-left panel, overlaid with magnetic field lines. 
	 The relativistic energy of Particle 2 is plotted versus time linear scale in the middle-left panel, where Phases 1, 2, and 3 are indicated with a green, yellow, or white background respectively. 
 The successively zoomed-in views of the energy curve are shown in the bottom-left panel, bottom-right panel and middle-right panel. The projected trajectory in the {\it x-y} plane during one cycle of bouncing motion of the particle is plotted in orange curve in the top-right panel, where the start (in white) and end (in black) positions of the trajectory are indicated with colored dots. The start time ($t_1$) and end time ($t_2$) of the motion are also annotated.   
	\label{energycurve}}	
\end{figure*}

The motion of proton 2 is divided into 3 phases. The duration of each phase is listed in the second column of Table~\ref{particlephase}. From top to bottom, the time variations of the relativistic energy, the gyroradius and the magnetic flux enclosed by orbits of gyration are plotted, respectively, in log scale in the left column of Fig.~\ref{single}. The bottom-right panel of Fig.~\ref{single} is a zoomed-in view of the magnetic flux enclosed by the gyro-motion orbits during Phase 1 and 2. Phases 1, 2, and 3 are indicated by a green, yellow, or white background respectively in the left column and in the bottom-right panel of Fig.~\ref{single}. 

The particle trajectory during Phases 1 and 2 are plotted in red in the top-right and middle-right panels of Fig.~\ref{single}, respectively. In both cases we mark the start point (in white) and end point (in black) of the trajectory with a colored dot. For both panels, the spatial distribution of the mass density in the region $[-0.03L_{0},0.05L_{0}]\times[-0.03L_{0},0.05L_{0}]$ is shown in the background, overlaid with magnetic field lines. Phase 1 starts from the instant the particle is placed into the system, i.e., $t=17.6\,\mathrm{s}$, and ends once the particle enters the monster island at $t=18.6\,\mathrm{s}$ as shown in the top-right panel of Fig.~\ref{single}. Phase 2 starts at $t=18.6\,\mathrm{s}$ and ends at $t=27.7\,\mathrm{s}$. Phase 3 starts from $t=27.7\,\mathrm{s}$ immediately after the particle enters a small island located in the white box in the middle-right panel of Fig.~\ref{single}. 
To clearly show the small island inside the white box, we plot the same white box on the background density map with the magnetic field overlaid in the top-left panel of Fig.~\ref{energycurve}.
The electric field distribution inside the white box is plotted with the magnetic field overlaid in the top-right panel of Fig.~\ref{energycurve}, which is a zoomed-in view of the tiny island.

 As shown in the middle-left panel of Fig.~\ref{single}, the gyroradius of the particle in Phase 1, which is about $10^{-4}\,\mathrm{km}$, fluctuates with time. The gyroradius of the particle in Phase 2, which ranges from about $10^{-4}\,\mathrm{km}$ to about $10^{-2}\,\mathrm{km}$, is smooth and continuous rather than fluctuating. 
 The gyroradius of the particle in Phase 1 and 2 is much smaller than the length of its trajectory of hundreds of kilometers, which means that the particle trajectory consists of a series of cycles of gyromotion in Phase 1 and 2. The bottom panels of Fig.~\ref{single} show that the magnetic flux fluctuates rapidly in Phase 1 while it is smooth in Phase 2. The rapid fluctuation in Phase 1 indicates that the magnetic flux varies a lot from a cycle to the next cycle. Thus the magnetic flux is not considered to be conserved in Phase 1, and the motion of Phase 1 is non-adiabatic. The motion in Phase 1 is dominated by the guiding center drift.
 The drift motion is clearly shown by the trajectory of the particle in the upper-right panel of Fig.~\ref{single}, that is, the particle moves perpendicular to the magnetic field rather than along the magnetic field lines in Phase 1. The smoothness of the curves in Phase 2 indicates that the magnetic flux does not change too much from a cycle of motion to the next cycle and can be considered as conserved at least in short time scales. The short time scales here mean the time scales that are comparable to the period of gyromotion. As discussed in Section~\ref{Adiabatic}, when the field varies slowly, the adiabatic invariants are not strictly conserved but also slowly change. Thus it is acceptable that the long-term evolution of the magnetic flux changes with time. Here the long-term evolution means a time scale comparable to the time duration of Phase 2.

 \begin{figure*}[p]
    \includegraphics[scale=0.49]{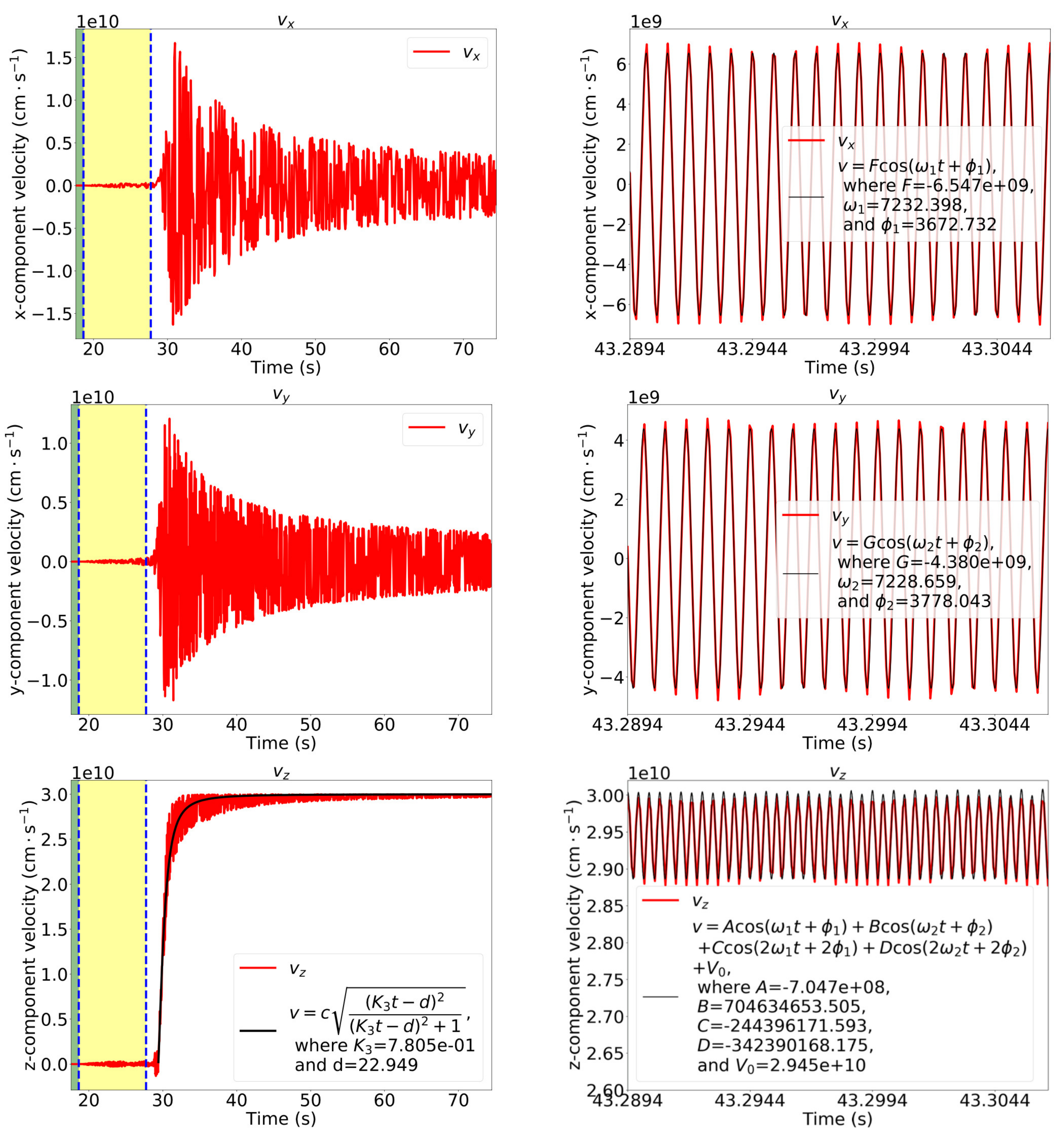}
	\caption{The three components of the velocity of Particle 2 are plotted by red curves in the left column. The periods of Phase 1 and Phase 2 are shaded by green and yellow respectively. The right column shows zoomed-in views  of the three components of the velocity of Particle 2 in the time period from $43.289\,\mathrm{s}$ to $43.307\,\mathrm{s}$. 
In right and bottom-left panels, the red curve is fitted with some function. The functions used to fit as well as the fitted parameters are annotated in each panel. The fitted functions are plotted by black curves.}\label{particleV}
\end{figure*}

 \begin{figure*}[p]
    \includegraphics[scale=0.35]{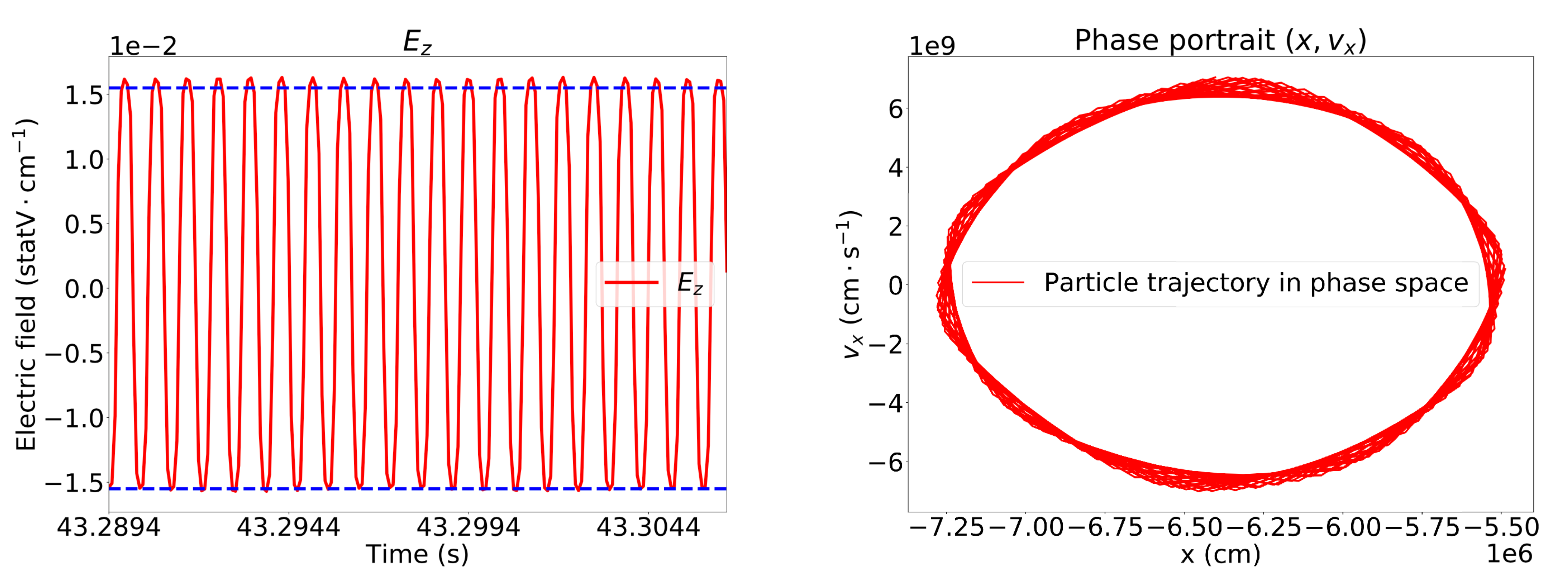} 
	\caption{ \label{Ephase}The left panel shows the variation of $E_{z}$ in the time period from $43.289\,\mathrm{s}$ to $43.307\,\mathrm{s}$. Two horizontal lines $E_{z}=\pm 1.55\times 10^{-2}\,\mathrm{statV\cdot cm^{-1}}$ are plotted, where $E_{z}=- 1.55\times 10^{-2}\,\mathrm{statV\cdot cm^{-1}}$ represents the valley of $E_{z}$. The phase portrait of the {\it x}-component coordinate and velocity of Particle 2 is plotted in the right panel.
	}
\vspace{1cm}
    \includegraphics[scale=0.45]{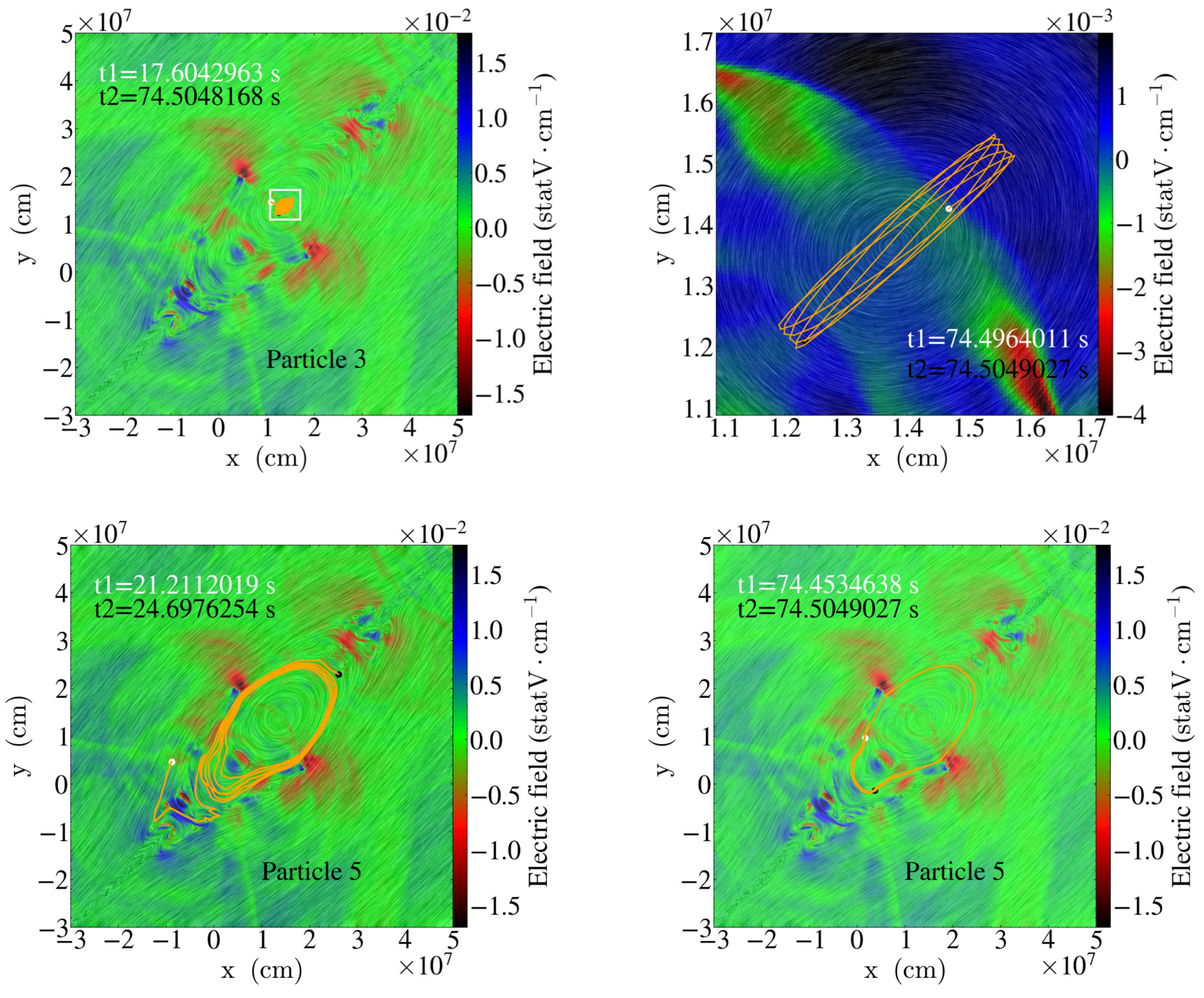}  
	\caption{\label{particle35} The trajectories of Particle 3 and 5 are plotted on the electric field background. Magnetic field lines are overlaid on each panel. The trajectories of Particle 3 are plotted in the top panels as orange curves in the time interval between $t_{1}$ and $t_{2}$ as annotated in each panel respectively. The top-right panel is a close-up view of the region in the white box with a different color scale in the top-left panel. The bottom panels show the trajectories of Particle 5 by orange curves in two time segments between $t_{1}$ and $t_{2}$ as annotated in each panel respectively. }
\end{figure*}

The energy of the particle is non-relativistic and changes little in Phase 1 and Phase 2 while the energy increases to hundreds of GeV within 30 s and becomes relativistic in Phase 3 as shown in the upper-left panel of Fig.~\ref{single}. Hence, Phase 3 is the main phase for particle acceleration. In order to understand the particle acceleration process, we analyze the motion of Phase 3 in detail.
The middle-left panel of Fig.~\ref{energycurve} shows the energy variation with time of the particle in linear scale, which is different from the log-scale plot in Fig.~\ref{energycurve} (bottom-right panel). The linear scale is used here in order to show the oscillations in the curve. Phases 1 and 2 are indicated with a green or yellow background, respectively. Phase 3 starts at $27.7\,\mathrm{s}$ when the particle enters the small island shown in the top-right panel. The energy starts to increase with time linearly from $t=30\,\mathrm{s}$. 
Subsequent zoom-ins into a small portion of the energy evolution during Phase 3 are shown in the bottom-left, bottom-right, and middle-right panels of Fig.~\ref{energycurve}. This shows that the energy evolution presents an oscillatory component in addition to a secular, linear growth. The period of the oscillation is about $8.6\times10^{-4}\,\mathrm{s}$. The projection of the trajectory of the particle in the {\it x-y} plane during one period of energy oscillation is plotted in the top-right panel of Fig.~\ref{energycurve} in orange curve, where the start and end positions of this cycle of oscillation are marked by the white and black dots respectively. The start and end instants of time are also shown in the plot. As the trajectory shows, the particle moves back and forth inside the island for many times and gains energy during each cycle.

 The three components of the velocity of the particle are plotted versus time in left panels of Fig.~\ref{particleV}. The right panels of Fig.~\ref{particleV} show the three components of the velocity during the same time period of the middle-right panels of Fig.~\ref{energycurve}. In right and bottom-left panels, the red curves represent the particle velocity components from the numerical simulation results while the black curves are the results of curve fitting by specific functions. The three components of the velocity oscillates with time, at different rates depending on the phase. The oscillation amplitudes of the three components are small during Phases 1 and 2 compared to Phase 3, during which they are relatively large. The amplitudes of the three components reach their peaks at $t=32\,\mathrm{s}$, after which the amplitudes start decreasing with time.
  The time-averaged values of the {\it x}- and {\it y}-components of the velocity are almost 0. The {\it z}-component velocity is the same in Phase 1 and 2 but it increases rapidly to the order of light speed in Phase 3. 
The {\it z}-component of the velocity during Phase 3 is fitted with the function
 \begin{equation}
 	v=c\sqrt{\frac{(K_{3}t-d)^{2}}{(K_{3}t-d)^{2}+1}},
 \end{equation} 
 where $c$ is the light speed, $K_{3}$ and $d$ are parameters to fit. The three components of the velocity in the time period considered in the right panels of Fig.~\ref{energycurve} are fitted with $v=F\cos(\omega_{1} t+\phi_{1})$, $v=G\cos(\omega_{2} t+\phi_{2})$ and $v=A\cos(\omega_{1} t+\phi_{1})+B\cos(\omega_{2} t+\phi_{2})+C\cos(2\omega_{1} t+2\phi_{1})+D\cos(2\omega_{2} t+2\phi_{2})+V_{0}$, respectively.
 The fitted functions are plotted as black curves in the corresponding panels. The reason to use the above functions to fit is explained in Section~\ref{solu0}.

Here we qualitatively analyze the reason why the particles are energized in Phase 3.
It is easy to verify from Equation (\ref{motion}) that the energy change rate of the particle in our two-dimensional setup is 
\begin{equation}
	\frac{\mathrm{d}}{\mathrm{d}t}\frac{mc^{2}}{\sqrt{1-(v/c)^{2}}}=qE_{z}v_{z}.
	\end{equation}
The energy change rate is fully determined by $E_{z}$ and $v_{z}$.
The variation of $E_{z}$ during the same time period of the middle-right panels of Fig.~\ref{energycurve} is plotted in the left panel of Fig.~\ref{Ephase}. Obviously, $E_{z}$ oscillates with time and changes its direction twice during each cycle of motion. The two horizontal dashed lines indicate the valley of the oscillation $-E_{V}$ and its absolute value $E_{V}=1.55\times 10^{-2}\,\mathrm{statV\cdot cm^{-1}}$ respectively. The peak value of the oscillation $E_{P}$ is larger than the absolute value of the valley.
The peak-to-valley ratio $\alpha_{PV}=E_{P}/E_{V}$ is larger than one. 
As shown in the bottom-left panel of Fig.~\ref{particleV}, $v_{z}$ is positive in the acceleration process in Phase 3. Meanwhile, the variation of $v_{z}$ during a cycle of motion $\delta v_{z}\sim 0.05 c$ is small compared to the value of $v_{z}\sim c$ as shown in the bottom-right panel of Fig.~\ref{particleV}. The energy $\Delta W$ gained by the particle during each cycle of motion can be qualitatively estimated as
\begin{equation}
	\Delta W \simeq q(-E_{V} + \alpha_{PV} E_{V})(v_{z}\pm \delta v_{z})\simeq (\alpha_{PV}-1)qE_{V}v_{z}.
\end{equation}  
If the peak-to-valley ratio $\alpha_{PV}$ is equal to 1, the particle will gain no energy. The particle gains energy because the peak of the electric field is larger than the absolute value of the valley.
Section~\ref{analytical} gives a more rigorous analysis of the statement.

\section{The motions of other particles}\label{otherparticle} 
The motion of Particle 1 is similar to the motion of Particle 2, going through the three phases of motion and trapped in the same small island as Particle 2. Particles 3 and 4 also go through the three phases of motion but are trapped around the center of the large monster magnetic island. The trajectory of Particle 3 is plotted in the upper-left panel of Fig.~\ref{particle35}. A small portion of this trajectory is shown in the top-right panel of the same figure. Particle 5 goes through Phase 1, then enters to a phase of adiabatic motion. The simulation stops at the phase of adiabatic motion and the non-adiabatic motion does not start. The energy of Particle 5 is almost conserved during the process as shown in bottom-right panel of Fig.~\ref{partinit}. Here we show the trajectories of Particle 5 in two time segments in the bottom panels of Fig.~\ref{particle35}. The bottom-left panel shows the trajectory at a relatively early stage of the adiabatic motion and the bottom-right panel shows the trajectory at the final stage of the simulation. 

 \section{Analytical investigation for the non-adiabatic motion of the particle around an O-point} \label{analytical}

An analytical model for the non-adiabatic motion of the particle constrained in a region smaller than its gyroradius around an O-point is constructed. The analytical model assumes that a two-dimensional magnetic O-point is located at the origin, and the particle motion in three-dimensional space around this O-point is investigated. The target of this analytical model is to explain the motion of particles during Phase 3 in the numerical simulations presented in the previous sections.

\subsection{Smoothing and decomposition}
 
The motion of a charged particle in an electromagnetic field is governed by Equation (\ref{motion}), where the electric field $\mathbf{E}(x,y)$ has only a {\it z}-component and the magnetic field $\mathbf{B}(x,y)$ lies in the {\it x-y} plane. Both fields are assumed to be stationary fields as functions of $x$ and $y$ only.
The only time-dependent variable is the velocity of the particle $\mathbf{v}(t)$.
The instantaneous velocity $\mathbf{v}(t)$ is decomposed into a time-averaged slowly changing part $\pmb{\mathcal{V}}(t)$ and a fast time-varying part $\pmb{\upsilon} (t)$, i.e.,
\begin{equation}
	\mathbf{v}(t)=\pmb{\mathcal{V}}(t)+ \pmb{\upsilon} (t)\label{vzdecompose}
\end{equation}  
or written in component form
\begin{equation}
	v_{i}(t)=\mathcal{V}_{i}(t)+ \upsilon_{i}(t)
\end{equation} 
where $i$ is taken as $x$, $y$ or $z$. 
The slow time-varying quantity $\pmb{\mathcal{V}}(t)$ is defined by a procedure of averaging $\mathbf{v}(t)$ over a time span $\Delta t$ at a moment of time $t$ as follows
\begin{equation}
	\pmb{\mathcal{V}}(t)=\frac{1}{\Delta t}\int^{t+\Delta t}_{t}\mathbf{v}(t^{\prime})\mathrm{d}t^{\prime}. \label{average}
\end{equation}
The fast time-varying term $\pmb{\upsilon} (t)$ is the deviation of $\mathbf{v}(t)$ from the slow time-varying term $\pmb{\mathcal{V}}(t)$, and its average is $\mathbf{0}$, i.e.,
\begin{equation}
	\frac{1}{\Delta t}\int^{t+\Delta t}_{t}\pmb{\upsilon}(t^{\prime})\mathrm{d}t^{\prime}=\mathbf{0}.
\end{equation}

The time interval $\Delta t$ is large enough such that the mean value $\pmb{\mathcal{V}}(t)$ does not change when $\Delta t$ increases. Meanwhile, $\Delta t$ is small in comparison with the characteristic time of the system evolution such that $\pmb{\mathcal{V}}(t)$ satisfies the condition of slowness of variation that $\pmb{\mathcal{V}}(t)$ varies little during the time interval $\Delta t$, i.e.,
 \begin{equation}
 \left|\frac{\mathrm{d} \pmb{\mathcal{V}}(t)}{\mathrm{d} t}\Delta t \right|\ll \left| \pmb{\mathcal{V}}(t)\right|. \label{slowness}
\end{equation}
The order of the time interval $\Delta t$ is about several periods of bouncing motion, close to the time span of the  right column of Fig.~\ref{particleV}. 
This decomposition is similar to the Reynolds decomposition in turbulence studies. In order to provide a more heuristic understanding of the decomposition, we refer to the bottom-left panel of Fig.~\ref{particleV}. The solid black curve obtained by data fitting represents the averaged variable $\mathcal{V}_{z}$ and the fluctuations on the red curve represent the fast time-varying variable $\upsilon_{z}$.  
The Lorentz factor $\gamma$ as a function of $t$ after $v$ can also be decomposed into a slow time-varying part $\Gamma(t)$ and a fast time-varying term $\mathfrak{r}(t)$, i.e.,
\begin{equation}
	\gamma(t)= \Gamma(t)+\mathfrak{r}(t),  \label{gammasplit}
\end{equation}
where
\begin{equation}
	\Gamma(t)=\frac{1}{\Delta t}\int^{t+\Delta t}_{t}\gamma(t)\mathrm{d}t  
\end{equation}
and 
\begin{equation}
	\mathfrak{r}(t)=\gamma(t)-\Gamma(t).   
\end{equation}
We derive the analytical expressions of $\Gamma(t)$ in the following Section.

\subsection{Model constraints and order analysis}\label{Modelconstraints}
We construct an analytical model that depicts the relativistic non-adiabatic motion of a charged particle around a two-dimensional magnetic O-point. Two constraints are assumed in this model, i.e., the range of the motion in the {\it x-y} plane should be smaller than the gyroradius and the long-term evolution of the particle should be relativistic. The first condition translates to  
\begin{equation}
	\mathcal{V}_{x}\ll \upsilon_{x}
\end{equation}
and
\begin{equation}
	\mathcal{V}_{y}\ll \upsilon_{y},
\end{equation}
meaning that the average speed in {\it x-y} plane is much smaller than the fluctuating velocity component, thus keeping the average displacement $L_{T}$ in {\it x-y} plane smaller than the gyroradius $r_{g}$. The particle is confined in a region of  a size smaller than the gyroradius $r_{g}$, otherwise if $\mathcal{V}_{x}\sim \upsilon_{x}\sim\mathcal{V}_{y}\sim \upsilon_{y}$, the average distance $L_{T}$ the particle moves during a cycle of motion will be comparable to the gyroradius, meaning that the particle escapes the confinement region that is smaller than the gyroradius during a cycle of motion, i.e.,
\begin{equation}
	L_{T}=\mathcal{V}_{y}\frac{2\pi}{\Omega_{g}}=\frac{2\pi\gamma mc\mathcal{V}_{y}}{qB}\sim r_{g}=\frac{\gamma mc\upsilon_{y}}{qB}.
\end{equation} 
The particle will move outside the confinement region after the evolution for a sufficiently long time if $\mathcal{V}_{x}$ or $\mathcal{V}_{y}$ are non-vanishing. So for a permanent confinement, i.e., the particle is permanently confined in a specific region, rather than a finite-time confinement, $\mathcal{V}_{x}$ and $\mathcal{V}_{y}$ have to be $0$. As shown in the top and middle panels of Fig.~\ref{particleV}, the average velocity components $\mathcal{V}_{x}$ and $\mathcal{V}_{y}$ are indeed almost zero. 
The condition that $\mathcal{V}_{x}\ll \upsilon_{x}$ and $\mathcal{V}_{y}\ll \upsilon_{y}$ implies that the slow time-varying speed $\mathcal{V}$ is dominated by $\mathcal{V}_{z}$. On account of the first constraint analyzed above, the condition that the long-term motion is relativistic requires that  
\begin{equation}
	\mathcal{V}\sim\mathcal{V}_{z}\sim c,
\end{equation}
which implies that
\begin{equation}
	\upsilon_{z}\ll \mathcal{V}_{z}.
\end{equation}
Obviously, the {\it z}-component of the velocity $v_{z}= (\upsilon_{z}+ \mathcal{V}_{z})$ may not exceed the light speed.
The above discussion shows that the following quantities can be regarded as small parameters
\begin{equation}
	\frac{\mathcal{V}_{x}}{\upsilon_{x}}\sim \frac{\mathcal{V}_{y}}{\upsilon_{y}} \sim \frac{\upsilon_{z}}{\mathcal{V}_{z}} \sim \frac{\upsilon }{\mathcal{V}}\ll 1. \label{smallpar}
\end{equation}
The numerical simulation results shown in Fig.~\ref{particleV} also illustrate the orders of magnitude of these quantities. As shown in Fig.~\ref{particleV}, the orders of magnitude of $v_{x}$, $v_{y}$ and $v_{z}$ are 
\begin{equation}
  v_{x}\sim v_{y}\sim\upsilon_{x}\sim \upsilon_{y}\sim 0.1c,
\end{equation} 
and 
\begin{equation}
	 v_{z}\sim\mathcal{V}_{z}\sim c,
\end{equation}
respectively. This ordering is indeed obtained for the particle we analysed in Phase 3 from Table~\ref{particlephase}.

In order to evaluate the orders of $\gamma(t)$, $\Gamma(t)$ and $\mathfrak{r}(t)$, we expand $\gamma$ in terms of $\upsilon/\mathcal{V}$
\begin{equation}
\begin{split}
	\gamma &=\frac{1}{\sqrt{1- (\pmb{\mathcal{V}}+\pmb{\upsilon} )^{2}/c^{2} }} \\&
	=\frac{1}{\sqrt{1- \mathcal{V}^{2}/c^{2} }}\left[1+\frac{1}{\left( 1- \mathcal{V}^{2}/c^{2} \right)}\frac{\mathcal{V}^{2}}{c^{2}}\left(\frac{ \pmb{\mathcal{V}}\cdot \pmb{\upsilon}}{\mathcal{V}^{2}}\right)+O\left(\frac{\upsilon^{2}}{\mathcal{V}^{2}}\right)\right],\label{gammaexpand}
\end{split}
\end{equation}
where $(\upsilon/\mathcal{V})$ is regarded as a small quantity as discussed above. Observe that the term containing $(\pmb{\mathcal{V}}\cdot \pmb{\upsilon}/\mathcal{V}^{2})$ vanishes after the average. Thus we have
\begin{equation}
	\Gamma(t)=\frac{1}{\Delta t}\int^{t+\Delta t}_{t}\frac{1}{\sqrt{1- \mathcal{V}^{2}/c^{2}}}\left[1+O\left(\frac{\upsilon^{2}}{\mathcal{V}^{2}}\right)\right]\mathrm{d}t 
\end{equation}
and 
\begin{equation}
\begin{split}
	\mathfrak{r}(t)&= \frac{1}{\left( 1- \mathcal{V}^{2}/c^{2} \right)^{3/2}}\frac{\mathcal{V}^{2}}{c^{2}}\left(\frac{ \pmb{\mathcal{V}}\cdot \pmb{\upsilon}}{\mathcal{V}^{2}}\right)\\&
	+\frac{1}{\sqrt{1- \mathcal{V}^{2}/c^{2}}}\left[O\left(\frac{\upsilon^{2}}{\mathcal{V}^{2}}\right)-\frac{1}{\Delta t}\int^{t+\Delta t}_{t}O\left(\frac{\upsilon^{2}}{\mathcal{V}^{2}}\right)\mathrm{d}t\right].
\end{split}
\end{equation}

\subsection{Approximation and correction}
Inequality (\ref{slowness}) plays an important role in the following analysis. We first limit our consideration to a short time interval $\Delta t$. The condition of slow variation of $\pmb{\mathcal{V}}(t)$ over $\Delta t$ implies that approximating $\pmb{\mathcal{V}}(t)$ as a constant $\pmb{\mathcal{V}}$ during $\Delta t$ is reasonable. Then the time dependence of the velocity $\mathbf{v}(t)$ solely comes from the fast time-varying part $\pmb{\upsilon}(t)$, i.e.,
\begin{equation}
	\mathbf{v}(t)=\pmb{\mathcal{V}}+ \pmb{\upsilon}(t).  \label{constplusfast}
\end{equation} 
We limit our consideration to permanent confinement, i.e., the particle keeps a finite distance to the O-point permanently. As pointed out in Section~\ref{Modelconstraints}, $\mathcal{V}_{x}$ and $\mathcal{V}_{y}$ have to be $0$ for permanent confinement, otherwise the particle will leave the O-point after a sufficiently long time. Then we have $\mathcal{V}=  \mathcal{V}_{z}$.
Inserting Equation (\ref{constplusfast}) into Equation (\ref{motion}), we obtain evolution equations for the fast time-varying variable $\pmb{\upsilon}(t)$ as follows
\begin{equation}
	\frac{\mathrm{d}(\gamma m \upsilon_{x})}{\mathrm{d}t}=-\frac{q}{c}B_{y}\mathcal{V}_{z}\left(1+\frac{\upsilon_{z}}{\mathcal{V}_{z}} \right), \label{1storder1}
\end{equation}

\begin{equation}
	\frac{\mathrm{d}(\gamma m \upsilon_{y})}{\mathrm{d}t}=\frac{q}{c}B_{x}\mathcal{V}_{z}\left(1+\frac{\upsilon_{z}}{\mathcal{V}_{z}} \right),             \label{1storder2}
\end{equation}
and 
\begin{equation}
\begin{split}
    \frac{\mathrm{d}(\gamma m \upsilon_{z})}{\mathrm{d}t}&=qE_{z}+\frac{q}{c}B_{y}\upsilon_{x}-\frac{q}{c}B_{x}\upsilon_{y}, \label{1storder3}
\end{split}
\end{equation}
where $\gamma(t)= \Gamma+\mathfrak{r}(t)$.

To first-order approximation, terms of order $(\upsilon/\mathcal{V})^2$ or higher should be dropped. The zeroth-order approximation can be formulated by dropping out terms of order $(\upsilon/\mathcal{V})$ or higher. We use subscripts to denote the orders of terms, e.g., $f_{0}$ represents the zeroth-order approximation for a quantity $f$, $f_{1}$ is the first-order correction, and $f_{2}$ is the second-order correction.

Dropping terms of order $(\upsilon/\mathcal{V})^2$ or higher in Equation (\ref{gammaexpand}), we obtain the first-order approximations of the slow time-varying part of $\gamma$ as
 \begin{equation}
    	\Gamma(t)=	 \Gamma_{0} +\Gamma_{1} =\frac{1}{\sqrt{1- \mathcal{V}_{z}^{2}/c^{2}}}  \label{gamma1storder1},
\end{equation}
and the fast time-varying part
\begin{equation}
	\mathfrak{r} (t)= \mathfrak{r}_{0} (t)+\mathfrak{r}_{1} (t)=\frac{1}{\left( 1- \mathcal{V}_{z}^{2}/c^{2} \right)^{3/2}}\frac{\mathcal{V}_{z}^{2}}{c^{2}}\left(\frac{\mathcal{V}_{z}\upsilon_{z}}{\mathcal{V}_{z}^{2}}\right),    \label{gamma1storder2}
\end{equation}
where we have taken $\mathcal{V}=\mathcal{V}_{z}$ by using the condition of permanent confinement $\mathcal{V}_{x}=\mathcal{V}_{y}=0$. As previously mentioned, $\mathcal{V}_{z}$ is regarded as a constant during $\Delta t$.
Here $\Gamma $ is a function of $\mathcal{V}_{z}$ only and thus can be regarded as a constant during $\Delta t$. 
The speed of the particle $v$ is expanded as follows
\begin{equation}
\begin{split}
		v&=\sqrt{\upsilon_{x}^{2}+\upsilon_{y}^{2}+(\mathcal{V}_{z}+\upsilon_{z})^{2}}\\&
		= \mathcal{V}_{z} \left[1+ \frac{  \upsilon_{z} }{\mathcal{V}_{z} }+O\left(\frac{\upsilon_{x}^{2}}{\mathcal{V}_{z}^{2}},\frac{\upsilon_{y}^{2}}{\mathcal{V}_{z}^{2}},\frac{\upsilon_{z}^{2}}{\mathcal{V}_{z}^{2}}\right)\right]\\&
		= \mathcal{V}_{z} \left[1+ \frac{  \upsilon_{z} }{\mathcal{V}_{z} }+O\left(\frac{\upsilon_{x}^{2}}{\mathcal{V}_{z}^{2}},\frac{\upsilon_{y}^{2}}{\mathcal{V}_{z}^{2}},\frac{\upsilon_{z}^{2}}{\mathcal{V}_{z}^{2}}\right)\right],
\end{split}
\end{equation}
where we have used the Taylor expansion formula $f(x)=(1+x)^{1/2}=1+(1/2)x+O(x^{2})$ of a function $f(x)$.
To first-order approximation, we have 
\begin{equation}
v =\mathcal{V}_{z} + \upsilon_{z}  .\label{speedorder}
\end{equation}
The zeroth-order approximation is obtained by dropping out those terms of the order $(\upsilon/\mathcal{V})$ or higher. Under the zeroth-order approximation, the fast time-varying part of $\gamma$ is taken as 0, i.e.,
 \begin{equation}
	\mathfrak{r}_{0}(t)= 0,
\end{equation}
the slow time-varying part of $\gamma$ is approximated as \begin{equation}
	\Gamma_{0}=\frac{1}{\sqrt{1- \mathcal{V}_{z}^{2}/c^{2}}}
\end{equation}
and the speed of the particle is
 \begin{equation}
	v_{0}=\mathcal{V}_{z}.\label{0orderv}
\end{equation}

The set of equations for the fast time-varying variable $\pmb{\upsilon} (t)$ under the zeroth-order approximation is 
\begin{equation}
	\Gamma_{0} \frac{ \mathrm{d}(m\upsilon_{x0})}{\mathrm{d}t}=-\frac{q}{c}B_{y}\mathcal{V}_{z} , \label{2ndorder1}
\end{equation}

\begin{equation}
	\Gamma_{0} \frac{ \mathrm{d}(m\upsilon_{y0})}{\mathrm{d}t}=\frac{q}{c}B_{x}\mathcal{V}_{z} ,  \label{2ndorder2}
\end{equation}
and 
\begin{equation}
\begin{split}
    \Gamma_{0} \frac{ \mathrm{d}(m\upsilon_{z0})}{\mathrm{d}t}&=qE_{z}+\frac{q}{c}B_{y}\upsilon_{x0}-\frac{q}{c}B_{x}\upsilon_{y_{0}},
    \label{2ndorder3}
\end{split}
\end{equation}
where $\Gamma_{0}$, treated as a constant during $\Delta t$, is taken out of the derivatives. It should be noted that our consideration is limited to permanent confinement, i.e. $\mathcal{V}_{x}=0$ and $\mathcal{V}_{y}=0$, and thus $v_{x}=\upsilon_{x}$ and $v_{y}=\upsilon_{y}$.

We first of all find out the zeroth-order solution $\pmb{\upsilon}_{0}$ by solving Equation (\ref{2ndorder1}) to (\ref{2ndorder3}). Then the first-order solution $\pmb{\upsilon}_{1}$ is found out by inserting $\pmb{\upsilon}=\pmb{\upsilon}_{0}+\pmb{\upsilon}_{1}$ into Equation (\ref{1storder1}) to (\ref{1storder3}) and using the first-order approximation of $\gamma$, i.e., Equation (\ref{gamma1storder1}) and (\ref{gamma1storder2}). The second-order solution $\pmb{\upsilon}_{2}$ is found out by iterating $\pmb{\upsilon}=\pmb{\upsilon}_{0}+\pmb{\upsilon}_{1}+\pmb{\upsilon}_{2}$ into Equation (\ref{1storder1}) to (\ref{1storder3}). Higher-order corrections can be obtained by the same iteration process.

\subsection{The magnetic configuration for the O-point}
\label{configuration_O}
In order to solve Equation (\ref{motion}), we need to specify the magnetic configuration for the two-dimensional O-point in which the particle moves. The two-dimensional magnetic field in the {\it x-y} plane is represented by a magnetic flux function $A(x,y)$ as
\begin{equation}
	\mathbf{B}=\nabla\times[A(x,y)\mathbf{e}_{z}].\label{BfromA}
\end{equation} 
To obtain the magnetic flux function at the O-point located at the origin, we expand $A(x,y)$ at the origin as follows
\begin{equation}
\begin{split}
	     A(x,y)&=A(0,0)+\frac{\partial A}{\partial x}\vert_{(0,0)}x+\frac{\partial A}{\partial y}\vert_{(0,0)}y\\&
	     +\frac{1}{2}\frac{\partial^{2} A}{\partial x^{2}}\vert_{(0,0)}x^{2}	
	     +\frac{1}{2}\frac{\partial^{2} A}{\partial y^{2}}\vert_{(0,0)}y^{2}
	     +\frac{\partial^{2} A}{\partial x\partial y}\vert_{(0,0)}xy\\&
	     +O(x^{3},y^{3}). \label{Aexp}
\end{split}
\end{equation}
Here $A(0,0)$ is a constant that can be dropped out. The O-point is a magnetic null point where the field vanishes, so that the first derivatives of the flux function are zero. We simply assume that the {\it x-y} axes coincide with the elliptical symmetry axes so that the term $[(\partial^{2} A/\partial x\partial y) xy]$ is crossed out. We limit our consideration to the second-order expansion at the current stage. Thus we have
\begin{equation}
	A=\frac{1}{2}ax^{2}+\frac{1}{2}by^{2}, \label{oflux}
\end{equation}
where $a= \partial^{2} A/\partial x^{2} $ and $b= \partial^{2} A/\partial y^{2} $. It has been proved (e.g., Chapter 1, Page 20 in Ref.~\onlinecite{PriestForbes2000}) that the above expression of $A$ represents the magnetic flux function around the O-point once $ab>0$. In Equation (\ref{oflux}), higher-order terms $O(x^{3},y^{3})$ are dropped out. The effects of these terms are discussed in Section \ref{distortion}. The magnetic field determined by Equation (\ref{oflux}) is 
\begin{equation}
	B_{x}=by,  \label{Bx}
\end{equation}
and
\begin{equation}
	B_{y}=-ax.  \label{By}
\end{equation}

\subsection{The electric field around the O-point}  
 \begin{figure*} 
    \includegraphics[scale=0.35]{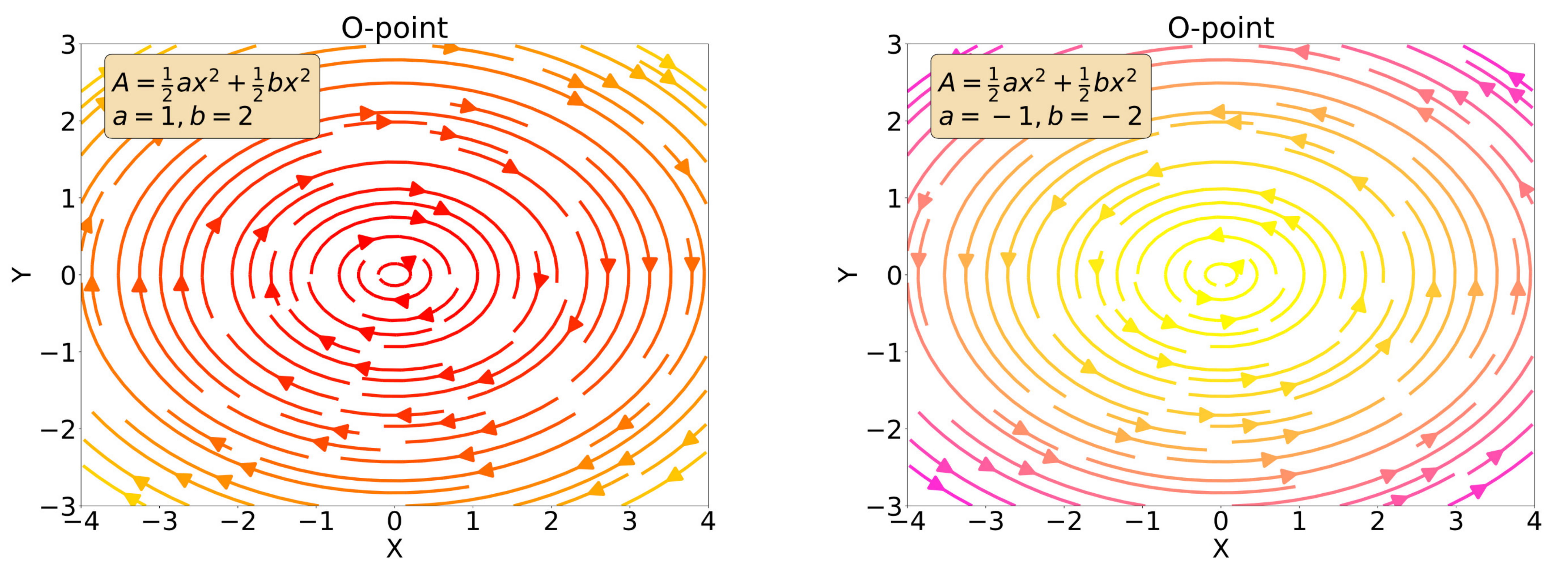}  
	\caption{The magnetic field lines around an O-point are represented by arrows in each panel. The magnetic field lines are clockwise when $a$ and $b$ are positive, and counterclockwise when $a$ and $b$ are negative.\label{Ostruct}}	
\end{figure*}
Now we consider the form of the electric field $\mathbf{E}$. The electric field consists of a resistive part $\eta \mathbf{j}$ and a convective part $-(1/c)\mathbf{u}\times \mathbf{B}$. As shown in Equation (\ref{Reynolds0}), the ratio of the convective term and the resistive term is the magnetic Reynolds number. In the case of high magnetic Reynolds numbers, the electric field $\mathbf{E}$ is dominated by the convective term $-(1/c)\mathbf{u}\times \mathbf{B}$, i.e.,
\begin{equation}
	E_{z}\simeq\frac{1}{c}(u_{y}B_{x}-u_{x}B_{y}). \label{generalelectric}
\end{equation}
We limit our discussion to the high-magnetic-Reynolds-number case, which is common in the solar corona.
We expand $u_{x}(x,y)$ and $u_{y}(x,y)$ at the origin as follows
\begin{equation}
\begin{split}
	     u_{i}(x,y)&=u_{i}(0,0)+\frac{\partial u_{i}}{\partial x}\vert_{(0,0)}x+\frac{\partial u_{i}}{\partial y}\vert_{(0,0)}y \\&
	          +O(x^{2},y^{2}),       \label{uexp}
\end{split}
\end{equation}
where $u_{i}$ represents either $u_{x}$ or $u_{y}$. 
To lowest-order approximation, we take $u_{i}(x,y) =u_{i}(0,0)=u_{i0}$ as constants. The effects of higher-order terms are discussed in Section~\ref{distortion}. Using the magnetic field given by Equations (\ref{Bx}) and (\ref{By}), we have
\begin{equation}
	E_{z}=\frac{1}{c}(bu_{y0}y+au_{x0}x). \label{0E}
\end{equation}

 \subsection{Solutions at the zeroth-order approximation\label{solu0}}
The set of equations for the fast time-varying variable $\pmb{\upsilon}(t)$ under the zeroth-order approximation is given by Equations (\ref{2ndorder1}) to (\ref{2ndorder3}). The electromagnetic field is given by Equations (\ref{Bx}), (\ref{By}) and (\ref{0E}). To find the zeroth-order solution $\pmb{\upsilon}_{0}$, we note that Equations (\ref{2ndorder1}) and (\ref{2ndorder2}) are independent of each other as $\mathcal{V}_{z}$ is treated as a constant during $\Delta t$. Inserting $\upsilon_{x0}$ and $\upsilon_{y0}$, which are the solutions of Equations (\ref{2ndorder1}) and (\ref{2ndorder2}) respectively, into Equation (\ref{2ndorder3}), we obtain the zeroth-order solution of Equation (\ref{2ndorder3}), which is denoted as $\upsilon_{z0}$.

The equation of motion of the particle in the {\it x}-direction under the zeroth-order approximation is
 \begin{equation}
   \Gamma m \frac{\mathrm{d}^{2}x}{\mathrm{d}t^{2}}=\frac{q}{c}a\mathcal{V}_{z}x 	\label{0ordereq}
 \end{equation}
 where $\Gamma$ and $\mathcal{V}_{z}$ are treated as constants.
In order to find the solution of Equation (\ref{0ordereq}),
a trial function $x=K\exp(\lambda t)$ is inserted and we immediately obtain
 \begin{equation}
 	\lambda^{2} =\frac{qa\mathcal{V}_{z}}{\Gamma mc} .
 \end{equation} 
 The solutions of Equation (\ref{0ordereq}) are divided into two classes, which is directly related to the sign of $a$ for our case of a proton with positive $\mathcal{V}_{z}$.
 
The first class of solution is obtained when $qa\mathcal{V}_{z}/(\Gamma c)>0$ as follows
  \begin{equation}
    x =K_{0}\exp\left(\sqrt{\frac{qa\mathcal{V}_{z}}{\Gamma mc}}t\right)-K_{0}\exp\left(-\sqrt{\frac{qa\mathcal{V}_{z}}{\Gamma mc}}t\right),
 \end{equation}
 where we have assumed that $x=0$ when $t=0$, i.e., the particle is located at the origin initially. The constant $K_{0}$ is determined by the initial velocity, and even a tiny initial velocity gives rise to a non-vanishing $K_{0}$, and the particle goes to infinity as time goes on. The sign of $qa\mathcal{V}_{z}/(\Gamma mc)$ is solely determined by $a$ under these conditions. The positive $a$ corresponds to a clockwise rotating magnetic field around the O-point as shown in the left panel of Fig.~\ref{Ostruct}. We thus reach a conclusion that an O-point with a clockwise rotating magnetic field around it can neither trap a positive charge with a positive {\it z}-component of the velocity nor a negative charge with a negative {\it z}-component of the velocity. In such a condition, the particle can not be confined around the O-point but can be scattered away from the O-point.

 The second class of solution is obtained when $qa\mathcal{V}_{z}/(\Gamma mc)<0$, which implies that $a<0$ and the magnetic field rotates counterclockwise around the O-point as shown in the right panel of Fig.~\ref{Ostruct}. The solution for this case is as follows
   \begin{equation}
    x =K_{1}\sin\left(\sqrt{\left|  \frac{qa\mathcal{V}_{z}}{\Gamma mc}\right|} t+\phi_{1}\right). \label{xsol}
 \end{equation} 
 Immediately, the velocity of the particle is obtained by taking the time derivative of $x(t)$
    \begin{equation}
    \upsilon_{x} =K_{1}\sqrt{\left|  \frac{qa\mathcal{V}_{z}}{\Gamma mc}\right|} \cos\left(\sqrt{\left|  \frac{qa\mathcal{V}_{z}}{\Gamma mc}\right|}  t+\phi_{1}\right).\label{vxsol}
 \end{equation}
 Applying the above analysis to the motion in the {\it y}-direction, we have
   \begin{equation}
    y =K_{2}\sin\left(\sqrt{\left|  \frac{qb\mathcal{V}_{z}}{\Gamma mc}\right|} t+\phi_{2}\right),\label{ysol}
 \end{equation}
 and 
\begin{equation}
    \upsilon_{y} =K_{2}\sqrt{\left|  \frac{qb\mathcal{V}_{z}}{\Gamma mc}\right|} \cos\left(\sqrt{\left|  \frac{qb\mathcal{V}_{z}}{\Gamma mc}\right|}t+\phi_{2}\right).\label{vysol}
\end{equation}  
The above analysis shows that the trajectories of the particle in the {\it x-y} plane are Lissajous curves. The motion of the particle in the {\it x}-direction and {\it y}-direction are uncoupled with each other and are analogous to harmonic oscillators with Hamiltonians of the forms  
\begin{equation}
	H_{x}=\frac{p_{x}^{2}}{2m} -\frac{qa\mathcal{V}_{z}}{2c}x^{2} 
\end{equation}
and
\begin{equation}
	H_{y}=\frac{p_{y}^{2}}{2m} -\frac{qb\mathcal{V}_{z}}{2c}y^{2} 
\end{equation}
in the {\it x-} and {\it y}-directions respectively with $p_{x}=\Gamma m(\mathrm{d}x/\mathrm{d}t)$ and $p_{y}=\Gamma m(\mathrm{d}y/\mathrm{d}t)$.

\begin{figure}[ht!]
    \includegraphics[scale=0.27]{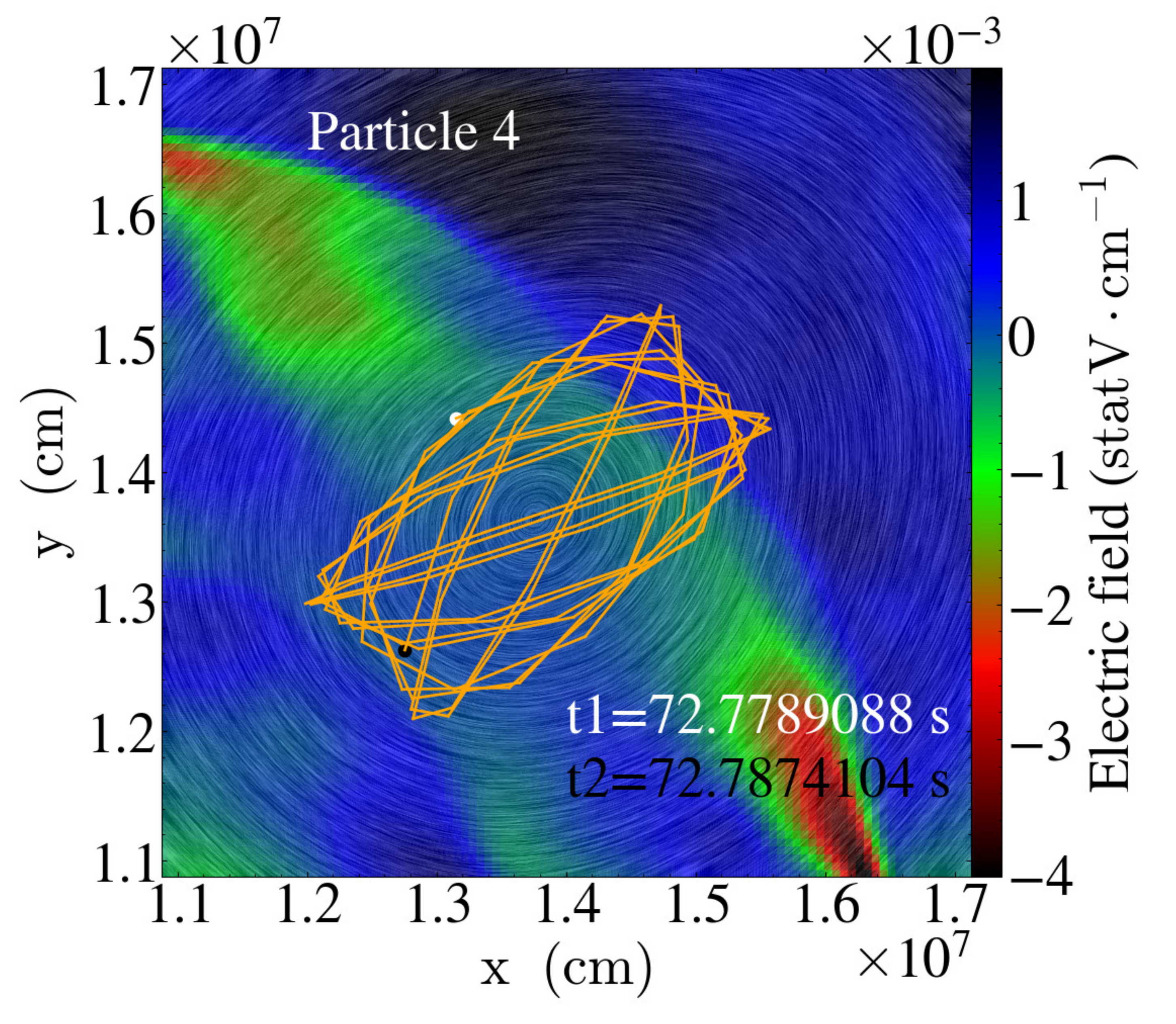}
	\caption{ The trajectory of Particle 4 is plotted on the electric field background. The magnetic field is overlaid on each panel by the LIC technique. The start and end positions are marked as white and black dots respectively. }\label{particle4}
\end{figure}

The orbit of the particle in the {\it x-y} plane is given by Equations (\ref{xsol}) and (\ref{ysol}). The orbit in the {\it x-y} plane is closed once the ratio of frequencies in the {\it x}-direction and {\it y}-direction is a rational number, i.e.,
\begin{equation}
	\frac{\sqrt{\mid  qa\mathcal{V}_{z}/(\Gamma mc)\mid}}{\sqrt{\mid  qb\mathcal{V}_{z}/(\Gamma mc)\mid}}=\frac{\sqrt{\mid a\mid}}{\sqrt{\mid b\mid}}=\mathrm{ rational}. \label{rational}
\end{equation}
This shows that whether the orbit in the {\it x-y} plane is closed is purely determined by $\sqrt{\mid a\mid}$ and $\sqrt{\mid b\mid}$, the two parameters determining the magnetic configuration of the O-point. 
Indeed, we find that the orbit of the proton in our counterclockwise island topology is a Lissajous curve, such that the particle stays trapped for a significant time. The phase trajectory in the {\it x}-direction determined by Equations (\ref{xsol}) and (\ref{vxsol}) will not be perfectly closed in general (or only in rare islands obeying Equation (\ref{rational})), but does follow an elliptic curve in phase space, as indeed seen in Figure \ref{Ephase}. As mentioned in Section~\ref{otherparticle}, Particles 3 and 4 are trapped around the center of the monster island, and their trajectories are Lissajous curves. A small portion of the trajectory of Particle 3 is shown in the top-right panel of Fig.~\ref{particle35}. The trajectory of Particle 4 in the time interval between $t_1$ and $t_2$ is shown as an orange curve in Fig.~\ref{particle4}.

\begin{figure}[ht!]
    \includegraphics[scale=0.2]{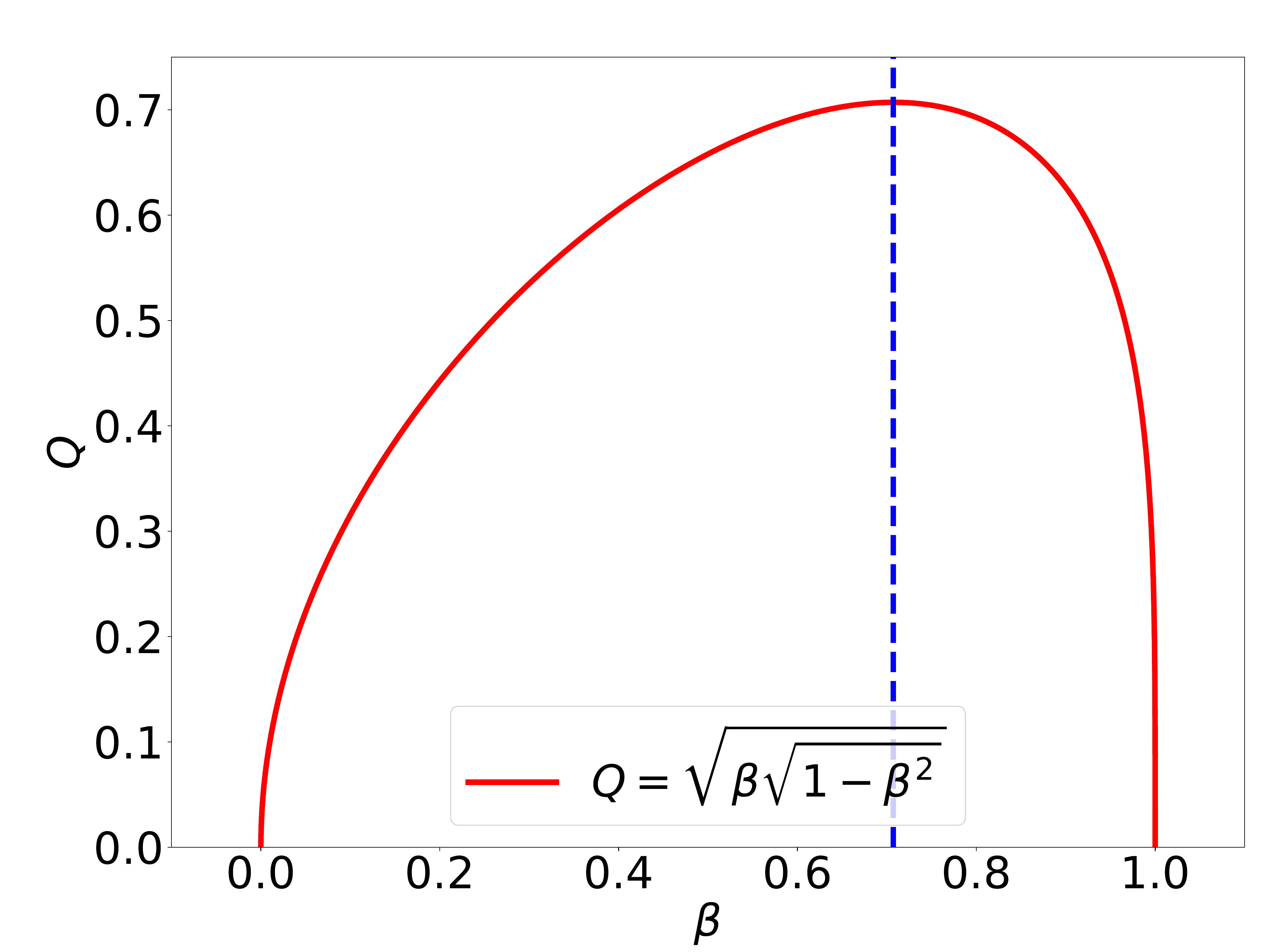}
	\caption{The function $Q= \sqrt{\beta \sqrt{1-\beta^{2} }}, 0<\beta<1$. The maximum occurs at $\beta=\sqrt{2}/2$, which is represented by the vertical dashed line.\label{function}}
\end{figure}

The short time interval $\Delta t$ in the averaging procedure in Equation (\ref{average}) should be smaller than the global evolution time scale $t_{glob}$ but larger than 
 \begin{equation}
 \frac{2\pi}{\sqrt{\mid  qa\mathcal{V}_{z}/(\Gamma mc)\mid} }<\Delta t<t_{glob}. \label{smallt}
 \end{equation}
The horizontal range of the right column of Fig.~\ref{particleV} gives a heuristic illustration of how large $\Delta t$ is. 
As previously stated, the slowly changing part of the {\it z}-component of the velocity $\mathcal{V}_{z}(t)$ is considered as a constant $\mathcal{V}_{z}$ during a short time interval with the same order as $\Delta t$. For a long-time evolution, i.e., when time scales are much larger than $\Delta t$ and comparable to the global evolution time scale $t_{glob}$
   \begin{equation}
 t_{glob}\sim  t \gg \Delta t,
 \end{equation}
the slowly changing part of the {\it z}-component of the velocity is no longer a constant $\mathcal{V}_{z}$ but varies with time as a function $\mathcal{V}_{z}(t)$. The {\it x}-component of the velocity $v_{x}=\upsilon_{x}$ is given by Equation (\ref{vxsol}), which implies 
\begin{equation}
    v_{x}(t) \propto \sqrt{\left|  \frac{qa\mathcal{V}_{z}}{\Gamma mc}\right|}.
\end{equation}
Considering the long-time evolution and the time-variability of $\mathcal{V}_{z}(t)$, we have 
\begin{equation}
    v_{x}(t) \propto \sqrt{\left|  \frac{qa\mathcal{V}_{z}(t)}{\Gamma mc}\right|}
    \propto \sqrt{\left|  \frac{\mathcal{V}_{z}(t)}{ c}\right|\sqrt{1-\frac{\mathcal{V}_{z}^{2}(t)}{c^{2}}  }}. \label{ampvx} 
\end{equation}
We now study the properties of the function 
\begin{equation}
Q= \sqrt{\beta \sqrt{1-\beta^{2} }}, 0<\beta<1,  
\end{equation}
which is equivalent to the last term in Equation (\ref{ampvx}) by variable substitution $\beta=\mathcal{V}_{z}(t)/c$. 
The function is plotted in Fig.~\ref{function}. The maximum occurs at $\beta=\sqrt{2}/2\approx 0.7$. We thus conclude that the {\it x}-component of the velocity $v_{x}$ reaches its maximum when the {\it z}-component of the velocity $\mathcal{V}_{z}=0.7c$. This is qualitatively consistent with the numerical simulation result as shown in Fig.~\ref{particleV}. In Fig.~\ref{particleV}, the envelope of the red curve showing the {\it x}-component of the velocity $v_{x}$ reaches its peak when $\mathcal{V}_{z}=2.5\times 10^{10}\,\mathrm{cm\cdot s^{-1}}$, i.e. $\mathcal{V}_{z}=0.8c$, which is qualitatively consistent with the theoretical analysis. The function $Q= \sqrt{\beta \sqrt{1-\beta^{2} }}$ increases when $\beta<\sqrt{2}/2$ and decreases when $\beta>\sqrt{2}/2$, which is also consistent with the numerical simulation result as shown in Fig.~\ref{particleV}. The above analysis for $v_{x}$ also applies to $v_{y}$. There exists energy transfer between the {\it z}-direction and {\it x-y} plane. According to Equation (\ref{0orderv}), the zeroth-order speed $v=\mathcal{V}_{z}$, we thus conclude that the fast changing kinetic energy in the {\it x-y} plane reaches its maximum when the velocity of the particle is about $v=0.7c$.
 
The equation of motion of the particle in the {\it z}-direction under the zeroth-order approximation is
 \begin{equation}
\begin{split}
    \Gamma\frac{\mathrm{d}( m \upsilon_{z})}{\mathrm{d}t}&=qE_{z}-\frac{q}{c}ax\upsilon_{x}-\frac{q}{c}by\upsilon_{y}. \label{zdynamic}
\end{split}
\end{equation}
Inserting Equations (\ref{0E}), (\ref{xsol}), (\ref{ysol}), (\ref{vxsol}), and (\ref{vysol}) into Equation (\ref{zdynamic}), we can solve for $\upsilon_{z}$ for its zeroth-order approximation
\begin{equation}
\begin{split}
		\upsilon_{z}&=-\frac{q}{\Gamma mc}[\frac{K_{2}bu_{y0}}{\omega_{2}}\cos(\omega_{2}t+\phi_{2} )+\frac{K_{1}au_{x0}}{\omega_{1}}\cos(\omega_{1}t+\phi_{1} ) ]\\&
		+\frac{q}{\Gamma mc}[\frac{K_{1}^{2}a }{4 }\cos(2\omega_{1}t+2\phi_{1} )+\frac{K_{2}^{2}b }{4 }\cos(2\omega_{2}t+2\phi_{2} )], \label{vzsol}
\end{split}
\end{equation}
where we have denoted $\omega_{1}=\sqrt{\mid  qa\mathcal{V}_{z}/(\Gamma mc)\mid} $ and $\omega_{2}=\sqrt{\mid  qb\mathcal{V}_{z}/(\Gamma mc)\mid} $.
This confirms the shape of the fitting function used in Fig.~\ref{particleV} (bottom-right panel).

\subsection{Electric power under the zeroth-order approximation}
We turn to study how much energy the particle gains from the electromagnetic field. The amount of work the particle receives during a time interval $\Delta t$ is
\begin{equation}
\Delta W=q \int_{0}^{\Delta t}( \mathbf{E}+\frac{1}{c}\mathbf{v}\times\mathbf{B}  )\cdot \mathbf{v}\mathrm{d}t=q \int_{0}^{\Delta t} E_{z}v_{z}\mathrm{d}t,\label{longenergy}
\end{equation} 
where we have used the property that the electric field has only the {\it z}-component for a two-dimensional MHD simulation. 


The average power $\langle P\rangle$ during $\Delta t$ is calculated as follows
\begin{equation}
	\langle P\rangle =\frac{\Delta W}{\Delta t}=\frac{1}{\Delta t}\int_{0}^{\Delta t}q E_{z}v_{z}\mathrm{d}t,\label{longpower}
\end{equation}
where the power $P=E_{z}v_{z}$ can be rewritten as follows
\begin{equation}
	 P=E_{z}v_{z}= E_{z}\mathcal{V}_{z}\left(1+\frac{\upsilon_{z}}{\mathcal{V}_{z}}\right).
\end{equation}
To evaluate the order of $ P $, we expand $E_{z}$ and $\upsilon_{z}$ as follows
\begin{equation}
	E_{z}=E_{z0}+E_{z1}+E_{z2}+E_{z3}+\cdots ,
\end{equation}
and
\begin{equation}
	\upsilon_{z}=\upsilon_{z0}+\upsilon_{z1}+\upsilon_{z2}+\upsilon_{z3}+\cdots .
\end{equation}
Since $(\upsilon_{z}/\mathcal{V}_{z})$ is a first-order small parameter as shown by Equation (\ref{smallpar}), so is $(\upsilon_{z0}/\mathcal{V}_{z})$, while $(\upsilon_{z1}/\mathcal{V}_{z})$ is a second-order small parameter, and $(\upsilon_{z2}/\mathcal{V}_{z})$ is a third-order small parameter, and so on. Thus we have the zeroth-order approximation of $P$ as
\begin{equation}
	P_{0}=E_{z0}\mathcal{V}_{z}
\end{equation}
and the first-order correction $P_{1}$ as  
  \begin{equation}
 	 P_{1}= E_{z1}\upsilon_{z0}+ E_{z0}\upsilon_{z1}.
 \end{equation}
Here $E_{z0}=(1/c)(bu_{y0}y+au_{x0}x)$ is the unperturbed electric field as given by Equation (\ref{0E}). Higher-order terms of the electric field $E_{z1}$, $E_{z2}$ and $E_{z3}$, etc, come from the higher-order terms in Equations (\ref{Aexp}) and (\ref{uexp}) that are neglected in deriving Equation (\ref{0E}). And $\upsilon_{z0}(t)$ is given by Equation (\ref{vzsol}). The time interval $\Delta t$ is smaller than the long-term evolution time scale but is larger than the fast-changing time scale, so $\mathcal{V}_{z}(t)$ is considered as a constant $\mathcal{V}_{z}$ during $\Delta t$. 
 
We now calculate the zeroth-order $\langle P\rangle$, which is
\begin{equation}
	\langle P\rangle_{0}=\frac{1}{\Delta t}\int_{T}^{T+\Delta t}q E_{z0}\mathcal{V}_{z}\mathrm{d}t.
\end{equation}
Inserting Equation (\ref{0E}) into Equation (\ref{longpower}), we have
\begin{equation}
\langle P\rangle_{0}=\frac{\mathcal{V}_{z}q}{c\Delta t}\int_{T}^{T+\Delta t}   (bu_{y0}y_{0}+au_{x0}x_{0})\mathrm{d}t,\label{0power}
\end{equation} 
where $x_{0}(t)$ and $y_{0}(t)$ as functions of $t$ are given by Equations (\ref{xsol}) and (\ref{ysol}). We integrate Equation (\ref{0power}) to obtain 
\begin{equation}
\begin{split}
	\langle P\rangle_{0} &=\frac{K_{1}\mathcal{V}_{z}qau_{x0}}{\omega_{1}c\Delta t}  [\cos(\omega_{1}T+\phi_{1})-\cos(\omega_{1}\Delta t+\omega_{1}T+\phi_{1})]
		\\&
	+\frac{K_{2}\mathcal{V}_{z}qbu_{y0}}{\omega_{2}c\Delta t}  [\cos(\omega_{2}T+\phi_{2})-\cos(\omega_{2}\Delta t+\omega_{2}T+\phi_{2})],\label{0powerresul}
\end{split} 
\end{equation} 
 where we have denoted $\omega_{1}=\sqrt{\mid  qav_{z0}/(\gamma mc)\mid} $ and $\omega_{2}=\sqrt{\mid  qbv_{z0}/(\gamma mc)\mid} $. In the limit $\Delta t\to \infty$, we find that
 \begin{equation}
 	\lim_{\Delta t\to \infty} \langle P\rangle_{0}=0.
 \end{equation}
 This limit is relevant because we consider $\Delta t$ to cover many fast time-varying fluctuations. The time for $\mathcal{V}_z(t)$ variations is much longer due to Equation (\ref{slowness}). The above analysis shows that the particle gains no energy from the electromagnetic field under the zeroth-order approximation.

 In order to understand the particle acceleration process, we have to consider first-order corrections. Under the first-order approximation, the average power $\langle P\rangle$ has the following form
  \begin{equation}
  \begin{split}
  \langle P\rangle &=\langle P\rangle_{0}\\&
  +\frac{1}{\Delta t}\int_{T}^{T+\Delta t}qE_{z0}\upsilon_{z0}\mathrm{d}t
  +\frac{1}{\Delta t}\int_{T}^{T+\Delta t}qE_{z1}\mathcal{V}_{z}\mathrm{d}t,\label{1power}
  \end{split}
 \end{equation}
 where we use $\langle P\rangle_{0}$ to denote the average power under the zeroth order approximation, which is given in Equation (\ref{0powerresul}).  
It is easy to verify that 
 \begin{equation}
 	\lim_{\Delta t \to \infty}\frac{1}{\Delta t}\int_{T}^{T+\Delta t}qE_{z0}\upsilon_{z0}\mathrm{d}t=0
 \end{equation}
 by inserting the expression of $\upsilon_{z0}$ given by Equation (\ref{vzsol}).
 
 We thus show that the particle can only gain energy from the third term on the right-hand side of Equation (\ref{1power}).

\subsection{The distortion of magnetic and velocity fields}\label{distortion} 
To calculate the third term on the right-hand side of Equation (\ref{1power}), we need to evaluate $E_{z1}$, i.e., the first-order correction to the electric field. The general expression of the {\it z}-component of the electric field is given by Equation (\ref{generalelectric}). In the zeroth-order approximation, we have taken $u_{i}=u_{i0}$ in Equation (\ref{uexp}) and the magnetic field is calculated from the magnetic flux function given by Equation (\ref{oflux}). Now we introduce the functions $\epsilon (x,y)$, $\varepsilon(x,y)$, and $\theta(x,y)$ as higher-order corrections to $u_{x}$, $u_{y}$ and $A$ respectively as follows
\begin{equation}
	u_{x}=u_{x0}\epsilon (x,y),\label{uxcorr}
\end{equation}
\begin{equation}
	u_{y}=u_{y0}\varepsilon(x,y),\label{uycorr}
\end{equation}
and 
\begin{equation}
	A=(\frac{1}{2}ax^{2}+\frac{1}{2}by^{2} )\theta(x,y),\label{Acorr}
\end{equation}
where $\epsilon (x,y)$, $\varepsilon(x,y)$, and $\theta(x,y)$ are equal to $1$ at the origin. We expand $\epsilon (x,y)$, $\varepsilon(x,y)$, and $\theta(x,y)$ as power series around the origin to first order in Equations (\ref{uxcorr}) to (\ref{Acorr}) as follows
\begin{equation}
	u_{x}=u_{x0}[1+\epsilon_{x}x+\epsilon_{y}y+O(x^{2},y^{2})],\label{uxcorr1} 
\end{equation}
		
\begin{equation}
	u_{y}=u_{y0}[1+\varepsilon_{x}x+\varepsilon_{y}y+O(x^{2},y^{2})],\label{uycorr1} 
\end{equation}
and 
\begin{equation}
	A=(\frac{1}{2}ax^{2}+\frac{1}{2}by^{2} )[1+\theta_{x}x+\theta_{y}y+O(x^{2},y^{2})],\label{Acorr1} 
	\end{equation}   
where $\epsilon_{x}$, $\epsilon_{y}$, $\varepsilon_{x}$, $\varepsilon_{y}$, $\theta_{x}$ and $\theta_{y}$ denote partial derivatives at the origin.   
Inserting the above into Equation (\ref{generalelectric}), we obtain the first-order correcting term $E_{z1}$ as follows
\begin{equation}
\begin{split}
		E_{z1}&=\left(\frac{\epsilon_{x}u_{x0}a}{c} + \frac{3\theta_{x}u_{x0}a}{2c}+\frac{\theta_{y}u_{y0}a}{2c}\right)x^{2}\\&
	+\left(\frac{\varepsilon_{y}u_{y0}b}{c}  	+\frac{3\theta_{y}u_{y0}b}{2c}+\frac{\theta_{x}u_{x0}b}{2c}\right)y^{2}\\&
	+\left(\frac{\varepsilon_{x}u_{y0}b}{c} +\frac{\epsilon_{y}u_{x0}a}{c}+\frac{\theta_{x}u_{y0}b}{c}+\frac{\theta_{y}u_{x0}a}{c}\right)xy. \label{Ez1}
\end{split}
\end{equation} 
Noting that the coefficients of $x^{2}$, $y^{2}$ and $xy$ are all constants, we denote
\begin{equation}
		E_{z1} =\kappa x^{2}+\sigma y^{2}+  \tau xy \label{kappasigma}
\end{equation} 
where $\kappa$, $\sigma$ and $\tau$ are the corresponding coefficients of $x^{2}$, $y^{2}$ and $xy$ respectively.
 This "perturbed" electric field is due to the variation of the plasma flow throughout the island, in combination with the island deviation from a perfectly elliptical shape. This is again something entirely known from the full MHD simulation in which we carry out our test-particle analyses. 

By using Equations (\ref{xsol}) and (\ref{ysol}), we thus obtain the result of the long-term average power under the first-order approximation
 \begin{equation}
 \begin{split}
 	\lim_{\Delta t\to \infty}\langle P\rangle &= \lim_{\Delta t\to \infty} \frac{1}{\Delta t}\int_{T}^{T+\Delta t}qE_{z1}\mathcal{V}_{z}\mathrm{d}t \\&=q\mathcal{V}_{z}\left(\frac{\kappa K_{1}^{2}}{2}+\frac{\sigma K_{2}^{2}}{2}\right). \label{finalpower}
 \end{split}
 \end{equation}
Noting that 
\begin{equation}
\lim_{\Delta t\to \infty}\langle P\rangle \sim	\kappa \sim \sigma \sim O\left(\frac{u}{c}\right),
\end{equation}
where $u$ is the order of magnitude of the fluid speed,
we reach the conclusion that the energy gained by the particle from the electromagnetic field during a cycle of motion with a period of $T_{period}$ is proportional to $u/c$
\begin{equation}
\Delta W =T_{period} \lim_{\Delta t\to \infty}\langle P\rangle  \sim O\left(\frac{u}{c}\right),
\end{equation} 
 which is similar to the first-order Fermi-type acceleration in the sense that the energy gain is proportional to the first power of $u/c$.

The slowly changing part of the {\it z}-component of the velocity can be obtained by integrating Equation (\ref{finalpower}) by noting that
\begin{equation}
	\lim_{\Delta t\to \infty}\langle P\rangle=\frac{\mathrm{d}}{\mathrm{d}t}\frac{mc^{2}}{\sqrt{1-\mathcal{V}_{z}^{2}(t)/c^{2}}}.
\end{equation}
Then $\mathcal{V}_{z}$ satisfies the following differential equation
\begin{equation}
	\frac{\mathrm{d}}{\mathrm{d}t}\frac{mc^{2}}{\sqrt{1-\mathcal{V}^{2}_{z}(t)/c^{2}}}=q\mathcal{V}_{z}(t)\left(\frac{\kappa K_{1}^{2}}{2}+\frac{\sigma K_{2}^{2}}{2}\right),
\end{equation}  
the solution of which is
\begin{equation}
	\mathcal{V}_{z}(t)=c\sqrt{\frac{(K_{3}t-d)^{2}}{(K_{3}t-d)^{2}+1}},
\end{equation}
where $d$ is a constant and  
\begin{equation}
	K_{3}=\frac{q}{mc} \left(\frac{\kappa K_{1}^{2}}{2}+\frac{\sigma K_{2}^{2}}{2}\right).\label{K_3}
\end{equation}
Here we note that the dimension of $[(\kappa K_{1}^{2}/2)+(\sigma K_{2}^{2}/2)]$ is the same as that of electric or magnetic fields under Gaussian units and $K_{3}$ has the dimension of frequency. The curve fitting in Section~\ref{particle_motion} shows that $K_{3}=0.7805\,\mathrm{s^{-1}}$ for the motion of Particle 2 during Phase 3. The long-term motion in the {\it z}-direction is equivalent to the motion of a charged particle in a constant uniform electric field directed along the {\it z}-axis. The effective electric field is
\begin{equation}
	\mathbf{E}_{eff}=\frac{mc}{q}K_{3}\mathbf{e}_{z} .
\end{equation}
The Hamiltonian of the motion in the {\it z}-direction is
\begin{equation}
	H_{z}=\frac{p_{z}^{2}}{2m}-mcK_{3}z,
\end{equation}
where $p_{z}=\gamma m(\mathrm{d}z/\mathrm{d}t)$.
The time required for a particle to be accelerated from zero velocity to $\alpha c$ is
\begin{equation}
	\tau=\frac{\alpha}{\sqrt{1-\alpha^{2}}}\frac{1}{K_{3}}
\end{equation}
where $0< \alpha <1$. For example, an initially static particle can reach a speed of $87\% $ of the light speed within $1.35\,\mathrm{s}$ by taking $K_{3}=0.7805\,\mathrm{s^{-1}}$.

\subsection{High-efficiency acceleration of particles at $0.7c$}
 
As discussed in Section~\ref{solu0}, the {\it x}- and {\it y}-components of the velocity $v_{x}$ and $v_{y}$ reach their maximum amplitude when the {\it z}-component of the velocity $\mathcal{V}_{z}=\sqrt{2}c/2$. The frequency of the particle motion in the {\it x}-direction $\omega_{1}$, which is given by Equation (\ref{xsol}), also reaches a maximum when $\mathcal{V}_{z}=\sqrt{2}c/2$ because 
 \begin{equation}
    \omega_{1}=\sqrt{\left|  \frac{qa\mathcal{V}_{z}}{\Gamma mc}\right|}  \propto \sqrt{\left|  \frac{\mathcal{V}_{z}(t)}{ c}\right|\sqrt{1-\frac{\mathcal{V}_{z}^{2}(t)}{c^{2}}  }},
 \end{equation} 
which is the same as the amplitude shown by Equation (\ref{ampvx}). Thus the bouncing motion of particles in the {\it x-y} plane reaches the highest frequency and speed when $\mathcal{V}_{z}=\sqrt{2}c/2$. As discussed in Section~\ref{distortion}, the amount of work a particle receives over each cycle of motion is proportional to $u/c$. So the higher the frequency of the back and forth movement in the x-y plane, the more efficient its acceleration.
Equation (\ref{speedorder}) shows the velocity of a particle is taken as $v=\mathcal{V}_{z}$ under the zeroth-order approximation. Then we reach the conclusion that particles with speeds close to $\sqrt{2}c/2\approx0.7c$ are accelerated with the highest efficiency.

 \subsection{Comparison between numerical and analytical results}  
 To compare the numerical and analytical results, we need to obtain the geometric parameters characterizing the shape of the magnetic and fluid velocity fields, i.e., $a$, $b$, $\epsilon$, $\varepsilon$, and $\theta$.
 
\begin{figure}[ht!]
    \includegraphics[scale=0.6]{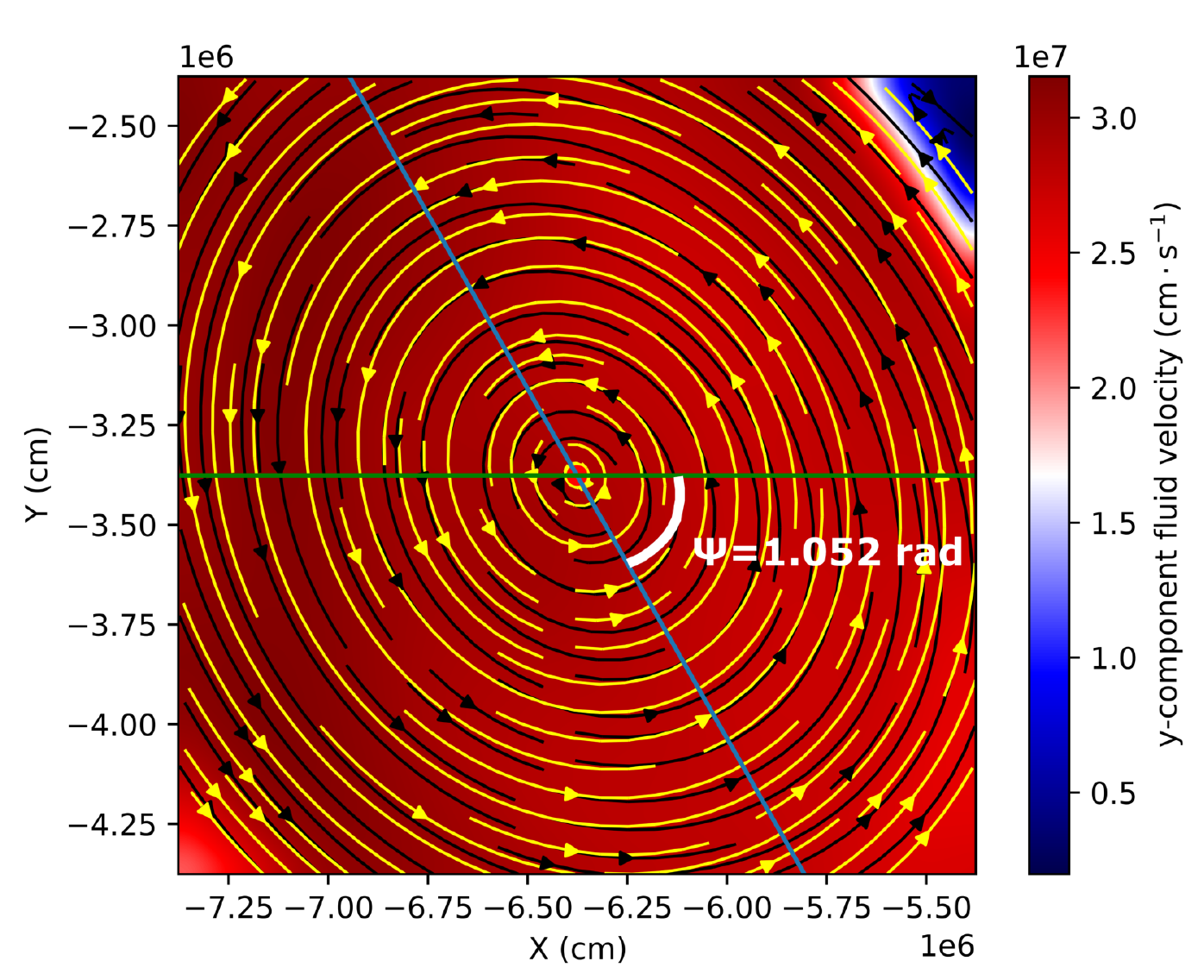}
	\caption{The {\it y}-component of the fluid velocity distribution with the simulated magnetic field (black arrows) and the fitted magnetic field (yellow arrows) overlaid. The green line represents the {\it x}-axis of {\it x-o-y} coordinate system while the blue line represents the $x^{\prime}$-axis of $x^{\prime}$-o-$y^{\prime}$ coordinate system, and the angle subtended by the two axes is $\psi=1.052\,\mathrm{rad}$.\label{fluidy}}
\end{figure}

The magnetic island in our simulation is not exactly located at the origin of the coordinate system. 
Now we pick the O-point in the center of the island where the magnetic field vanishes and shift the origin of the coordinate system to the O-point. In this new coordinate system $x$-o-$y$, the magnetic flux function approximated to second order around the O-point has the following form
\begin{equation}
\begin{split}
	A=	\frac{1}{2}a_{0}x^{2}  
	  + \frac{1}{2}b_{0}y^{2}+c_{0}xy, \label{rotateA}
\end{split}
\end{equation}
where $a_{0}$, $b_{0}$ and $c_{0}$ are constants.
The $xy$ term does not appear in Equation (\ref{oflux}) because it is dropped by placing the coordinate axes in alignment with the axes of the island.
However, in our simulation, the axes of the island are not well aligned with the coordinate axes, so we keep the $xy$ term in Equation (\ref{rotateA}). The relations between $a_{0}$, $b_{0}$ and $c_{0}$ in Equation (\ref{rotateA}) and $a$ and $b$ in Equation (\ref{oflux}) are as follows: $a_{0}=  a \cos^{2}\psi  +  b \sin^{2}\psi  $, $b_{0}= a \sin^{2}\psi + b \cos^{2}\psi $ and $c_{0}=-( a  - b  )\cos\psi \sin\psi $.  
Here a new parameter $\psi$ characterizing the angle between the major axis of the elliptical magnetic field lines and the {\it x}-axis is introduced. The meaning of $\psi$ can also be interpreted as follows: by rotating the coordinate system $x$-o-$y$ clockwise around the O-point through an angle $\psi$, we can obtain a new coordinate system $x^{\prime}$-o-$y^{\prime}$ where the $xy$ term vanishes and $A$ has the same form as Equation (\ref{oflux}). To illustrate the rotation of the coordinate system, we plot the {\it y}-component of the fluid velocity of the tiny island that traps Particle 2 in Fig.~\ref{fluidy}. The green line represents the {\it x}-axis of the $x$-o-$y$ coordinate system while the blue line represents the $x^{\prime}$-axis of $x^{\prime}$-o-$y^{\prime}$ coordinate system, and the angle subtended by the two axes is $\psi=1.052\,\mathrm{rad}$.

The magnetic field $\mathbf{B}$ determined via the magnetic flux function in Equation (\ref{rotateA}) is 
\begin{equation}
	B_x=b_{0}y+c_{0}x              \label{rotateB1}   
\end{equation} 
and
\begin{equation}
	B_y=-a_{0}x-c_{0}y.            \label{rotateB2}
\end{equation} 
 By fitting Equations (\ref{rotateB1}) and (\ref{rotateB2}) to the computed magnetic field, we obtain the values of $a_0$, $b_0$ and $c_0$, and thus obtain $a$, $b$ and $\psi$. 
The fitted values of the three parameters are $a_{0}=-2.32\times10^{-5}\,\mathrm{G\cdot cm^{-1}}$, $b_{0}=-1.98\times10^{-5}\,\mathrm{G\cdot cm^{-1}}$, and $c_{0}=-3.71\times10^{-6}\,\mathrm{G\cdot cm^{-1}}$.     
Then we can fit to obtain the parameter $\theta$ that characterizes the distortion of the magnetic field lines from the standard ellipse. By doing so, we write down Equation (\ref{Acorr1}) in $x$-o-$y$ as follows:
\begin{equation}
\begin{split}
		A=	&(\frac{1}{2}a_{0}x^{2}  
	  + \frac{1}{2}b_{0}y^{2}+c_{0}xy) 
	  [1+
	  \theta_{x}(x\cos\psi+y\sin\psi)\\&+\theta_{y}(-x\sin\psi+y\cos\psi) ]. \label{Acorr1rotate} 
\end{split}
\end{equation}   
By fitting the above expressions for the magnetic field to the computed data, we can obtain $\theta_{x}$ and $\theta_{y}$.
In the same way, the parameters characterizing the non-uniformity of the fluid velocity, $\epsilon$ and $\varepsilon$, can be obtained by fitting the following expressions to the computed data:
\begin{equation}
\begin{split}
	u_{x}=	&u_{x0}
	  [1+
	  \epsilon_{x}(x\cos\psi+y\sin\psi)\\&+\epsilon_{y}(-x\sin\psi+y\cos\psi) ] 
\end{split}
\end{equation} 
and
\begin{equation}
\begin{split}
		u_{y}=	&u_{y0}
	  [1+
	  \varepsilon_{x}(x\cos\psi+y\sin\psi)\\&+\varepsilon_{y}(-x\sin\psi+y\cos\psi) ].   
\end{split}
\end{equation} 
In Figure \ref{fluidy}, the black lines and arrows represent the magnetic field lines obtained from the numerical simulation while the yellow lines and arrows represent the fitted magnetic field lines obtained by fitting Equation (\ref{Acorr1rotate}) to the simulated data. The magnetic field lines are counterclockwise consistent with the discussion in Section~\ref{solu0}.

As mentioned in Section~\ref{otherparticle}, Particles 1 and 2 are trapped in a tiny island while Particles 3 and 4 are trapped around the center of a monster island. We list the fitted geometric parameters associated with the tiny island and the monster island in Table~\ref{islandparame}. The coefficients in Equation (\ref{kappasigma}), $\kappa$, $\sigma$ and $\tau$, are listed in Table~\ref{paramekappa}. 
As shown by Equation (\ref{rational}), the ratio of the frequencies in the {\it x} and {\it y}-directions are equal to $\sqrt{\mid a\mid/\mid b \mid}$. For Particle 2, the frequency ratio is $0.99$ while $\sqrt{\mid a\mid/\mid b \mid}=0.85$. The two values are not exactly the same but have the same order of magnitude. The parameter $K_{3}$ by the analytical prediction can be calculated according to Equation (\ref{K_3}) as $0.9109\,\mathrm{s}^{-1}$ while the fitted value is $0.7805\,\mathrm{s}^{-1}$.

\begin{table*} [ht!]
 	 
\caption{Geometric parameters of islands.\label{islandparame}}
         \resizebox{180mm}{6mm}
        {
\begin{tabular} {r|p{2.3cm}|r| p{0.9cm}| r| r| p{1.95cm}| r| r| r } 
\hline\hline
  &$a\,(10^{-5}\,\mathrm{G\cdot cm^{-1}})$ & $b\,(10^{-5}\,\mathrm{G\cdot cm^{-1}})$ & $\psi\,(\mathrm{rad})$ & $\theta_x\,(10^{-9} \,\mathrm{cm^{-1}})$ & $\theta_y \,(10^{-9} \,\mathrm{cm^{-1}})$ & $\epsilon_x\,(10^{-8}\,\mathrm{cm^{-1}})$ & $\epsilon_y\,(10^{-8}\,\mathrm{cm^{-1}})$& $\varepsilon_x\,(10^{-8}\,\mathrm{cm^{-1}})$& $\varepsilon_y\,(10^{-8}\,\mathrm{cm^{-1}})$\\\hline
Tiny island  & $-1.195 $ & $-1.636 $ & $1.052$ & $6.237 $& $21.14$ & $-9.480$& $-6.978$& $-10.24$& $-1.350$\\

monster island & $-0.4536$ & $-0.6560$ & $-0.684$ & $-5.804$& $34.16$ & $-0.4231$& $135.9$& $94.63$& $122.4$\\
\hline
\end{tabular}}
 
\end{table*}

\begin{table*} [ht!]
\begin{center} 
\caption{Coefficients of the first-order electric field.\label{paramekappa}}
        {
\begin{tabular}{r|r|r|r }
\hline\hline
  &$\kappa\,(10^{-16}\,\mathrm{statV\cdot cm^{-2}})$ & $\sigma\,(10^{-16}\,\mathrm{statV\cdot cm^{-2}})$ & $\tau\,(10^{-16}\,\mathrm{statV\cdot cm^{-2}})$  \\\hline
Tiny island  & $3.719 $ & $-3.121 $ & $21.99 $ \\

Monster island & $-0.3411$ & $1.770$ & $2.643$ \\
\hline
\end{tabular}}
 \end{center} 
\end{table*}

\section{Discussion}  
\label{discussion}

Our model demonstrates that particles can be energized in the vicinity around an O-point by non-adiabatic motion. However, the applications to actual solar or astrophysical scenarios are limited. Although there are limitations to our model, we can still gain much insight into acceleration of particles trapped in plasmoids. The main limitations and implications are discussed as follows.

\subsection{Diffusion versus advection \label{sechighmagneticReynolds}}
 
The analytical investigation for the non-adiabatic motion of the particle around an O-point in Section \ref{analytical} is based on the assumption that the magnetic Reynolds number is much larger than unity. Thus the electric field $E_{z}$ is dominated by the convective term $E_{con}=(1/c)(u_{y}B_{x}-u_{x}B_{y})$ and the resistive term $E_{res}=\eta J_{z}$ is negligible as given by Equation (\ref{generalelectric}). As shown by Equation (\ref{Reynolds}), the global magnetic Reynolds number $R_{m}^{G}$ is indeed much larger than unity in our MHD simulation. The global magnetic Reynolds number $R_{m}^{G}$ is calculated based on the typical global length scale $l=L_{0}$ and the characteristic reconnection inflow speed $u=0.1v_{0}$, which characterizes the global property of the system.
The high global magnetic Reynolds number implies that the electric field is dominated by the convective term in most of the simulation domain, except some boundary layers where the gradients of the magnetic field are steep. It is necessary to check whether the regions of our interest are those exceptions where the high-magnetic-Reynolds-number approximation breaks down. To do so, the local magnetic Reynolds number $R_{m}^{L}$ in the region of our interest should be calculated based on the local fluid speed and the length scale of interest. In our simulation, the non-adiabatic motion of the particle is confined in the vicinity of an O-point. The length scale of our interest is the range of the non-adiabatic motion in the {\it x-y} plane, which is about $10^{-3}L_{0}$ for Particle 2 as shown in the top-right panel of Fig.~\ref{energycurve}. The fluid speed in this region is about $2.5v_{0}$ as shown in Fig.~\ref{fluidy}, which is much larger than the reconnection inflow speed $0.1v_{0}$. Here the values of $L_{0}$ and $v_{0}$ are listed in Table~\ref{Units}. We thus obtain the local magnetic Reynolds number in the vicinity of the O-point in the tiny island as $R_{m}^{L}\simeq 2.5\times10^{3}$, which is much larger than unity. So the convective term $E_{con}$ is much larger than the resistive term $E_{res}$ in most of the region of our interest. However, our analysis in Section \ref{analytical} shows that the particle gains energy from the first-order correcting term $E_{z1}$ of the convective electric field. We need to compare the orders of $E_{res}$ and $E_{z1}$ rather than simply neglect $E_{res}$.  
If the resistive term $E_{res}$ is smaller than the first-order correcting term $E_{z1}$, the resistive term $E_{res}$ can be neglected. The first-order correcting term $E_{z1}$ of the electric field is given by Equation (\ref{kappasigma}), where the coefficients $\kappa$, $\sigma$, and $\tau$ are listed in Table~\ref{paramekappa}. By taking the length scale $l= 10^{-3}L_{0}$, we obtain the first-order correcting term $E_{z1}\simeq 2\times 10^{-3}\,\mathrm{statV\cdot cm^{-1}}$ for the tiny island. To estimate the resistive term $E_{res}=\eta J_{z}$, we should calculate the current density $J_{z}$ at first. The current density is given by 
\begin{equation}
J_{z}= [\frac{c}{4\pi}\nabla\times \nabla\times(A\mathbf{e}_{z})] \cdot\mathbf{e}_{z} =-\frac{c}{4\pi}\nabla^{2}A,	
\end{equation}
where $\mathbf{e}_{z}$ is the unit vector in the {\it z}-direction, and the magnetic flux function is given by Equation (\ref{Acorr1}). Equation (\ref{Acorr1}) can be written as
\begin{equation}
	A= \frac{1}{2}ax^{2}+\frac{1}{2}by^{2}+O(x^{3},y^{3}).  
	\end{equation} 
So we have  
\begin{equation}
J_{z}=-\frac{c}{4\pi}(a+b)+O(x,y)\simeq -\frac{c}{4\pi}(a+b),
\end{equation}
where the parameters $a$ and $b$ are listed in Table~\ref{islandparame}. The resistive term $E_{res}=\eta J_{z}$ for the tiny island is thus estimated as $E_{res}\simeq 10^{-5}\,\mathrm{statV\cdot cm^{-1}}$, which is much smaller than the first-order correcting term $E_{z1}\simeq 2\times 10^{-3}\,\mathrm{statV\cdot cm^{-1}}$. Hence we have verified that $E_{con}\gg E_{z1} \gg E_{res}$. The electric field in the vicinity of the O-point in the tiny island is dominated by the convective term $E_{con}$ and the resistive term $E_{res}$ is negligible. The above order analysis is quite consistent with the numerical simulation results. In our MHD simulation results, the order of magnitude of the resistive term $E_{res}$ is $10^{-5}\,\mathrm{statV\cdot cm^{-1}}$ while it is $10^{-2}\,\mathrm{statV\cdot cm^{-1}}$ for the convective term $E_{con}$, which gives a local magnetic Reynolds number $R_{m}^{L}\simeq 10^{3}$. The situation in the vicinity of the O-point in the monster island is similar to the tiny island.

The order analysis above shows that the order of the convective term $E_{con}$ is much larger than the order of the resistive term $E_{res}$ in most of the region with a length scale of $l=10^{-3}L_{0}$ around an O-point where the non-adiabatic motion occurs. However, there may exist some boundary layers or singular points in smaller scales than $l=10^{-3}L_{0}$ where high-magnetic-Reynolds number approximation breaks down and diffusion is important. It should be noted that the convective term $E_{con}=(1/c)(u_{y}B_{x}-u_{x}B_{y})$ is exactly zero at the O-point because the magnetic field vanishes at the null point. The resistive term $E_{res}=-(c/4\pi)(a+b)$ is non-zero at the O-point. Therefore, there exists a small area around the O-point where the resistive term $E_{res}$ dominates. The length scale of such an area is $l_{d}=4\times 10^{-7}L_{0}$, which is obtained by solving the equation $R^{L}_{m}=1$, i.e.,
\begin{equation}
	\frac{4\pi}{c^{2}}\frac{ul_{d}}{\eta}=1,
\end{equation} 
where $u=2.5v_{0}$ is the fluid speed around the O-point. We thus have
\begin{equation}
	\left\{\begin{matrix}
E_{res}\ge E_{con}, &  r_{o}\le l_{d}\\ 
E_{res}< E_{con}, &   l_{d}<r_{o}\lesssim 10^{-3}L_{0}
\end{matrix},\right.
\end{equation} 
where we denote the distance from any point in the {\it x-y} plane to the O-point in the tiny island as $r_{o}$. Noting that $l_{d}=4\times 10^{-7}L_{0}$ is much smaller than the smallest numerical grid size $6.1\times 10^{-5}L_{0}$, we find that the small region around the O-point where the resistive term $E_{res}$ dominates is unresolvable in our simulation. In this sense, the electric field $E_{z}$ is dominated by the convective term $E_{con}=(1/c)(u_{y}B_{x}-u_{x}B_{y})$ in the region where the non-adiabatic motion occurs and the resistive term $E_{res}=\eta J_{z}$ is negligible everywhere in this region. We thus verify that our analysis in Section \ref{analytical} is reasonable.

By solving the equation $R^{L}_{m}=1$, we can also estimate the thickness of the diffusion layer in the CS, which is between $10^{-5}L_{0}$ and $10^{-3}L_{0}$. The thickness of the diffusion layer is highly dependent on the reconnection inflow speed. The reconnection inflow speed varies with time and space in our simulation. In the region where the inflow speed is about $u=0.01v_{0}$, the thickness of the diffusion layer is $10^{-4}L_{0}$ while it is $10^{-5}L_{0}$ in the region with an inflow speed of $u=0.1v_{0}$. This is reasonable because faster inflow speed indicates a faster reconnection rate, which requires a steeper magnetic field gradient (thinner diffusion layer) to dissipate magnetic energy. The above crude estimation shows that the diffusion layer in the CS is covered by 1-100 numerical cells, which is consistent with our numerical simulation results.

\subsection{Adiabatic motion versus non-adiabatic motion} 
 
When the gyroradius $r_{g}$ of a particle is larger than or comparable with the curvature radius $r_{\kappa}$ of the magnetic field lines, i.e.,
\begin{equation}
	r_{\kappa}\lesssim r_{g}, \label{radiuscomp}
\end{equation}
the motion of the particle is considered to be non-adiabatic~\cite{Fu2006PhPl}.
 For an adiabatic motion, the curvature radius of the magnetic field lines is much larger than the gyroradius of the particle. We now discuss the adiabaticity of the motion in the vicinity of an O-point by comparing the gyroradius with the curvature radius of the magnetic field lines.

The relativistic gyroradius of a particle with mass $m$ and electric charge $q$ is
\begin{equation}
	r_g=\frac{mc^{2}}{|q|B}\sqrt{\gamma^{2}-1}.
\end{equation}
Here the magnitude of the magnetic field $B$ is proportional to $r_{o}$, which is easily seen from Equations (\ref{Bx}) and (\ref{By}), i.e.,
\begin{equation}
	B\sim B_{x}\sim B_{y} \sim by \sim ax \sim ar_{o}\sim br_{o},  
\end{equation}
where the values of $a$ and $b$ are listed in Table~\ref{islandparame}, and we use $r_{o}$ to denote the distance from a point located at $(x,y)$ to the O-point as in Section~\ref{sechighmagneticReynolds}. Here we take $a\sim b \sim 10^{-6}\,\mathrm{gauss\cdot cm^{-1}}$.  
The shape of the magnetic field lines around an O-point is an ellipse as discussed in Section~\ref{configuration_O}. Then the curvature radius of the magnetic field line at a point $(x,y)$ located in the vicinity of an O-point has the same order of $r_{o}$, i.e.,
\begin{equation}
	r_{\kappa}\sim r_{o}.
\end{equation} 
As discussed in Section~\ref{solu0}, particles at a speed of $\sqrt{2}/2c$ have the highest acceleration efficiency. So we take $v\simeq\sqrt{2}/2c$, which gives $\gamma=\sqrt{2}$. 
Thus Equation (\ref{radiuscomp}) is reduced to
\begin{equation}
	r_{o}\lesssim \frac{mc^{2}}{|q|ar_{o}} .
\end{equation} 
For electrons, we find
\begin{equation}
	r_{o}\lesssim 4\times10^{4}\,\mathrm{cm},
\end{equation} 
while it is 
\begin{equation}
	r_{o}\lesssim 2\times10^{6}\,\mathrm{cm},
\end{equation} 
 for protons.
 The above results show that the motion of a proton is non-adiabatic when its distance from the O-point is about $10^{6}\,\mathrm{cm}$. In our simulation, as shown in the top-right panel of Fig.~\ref{energycurve}, the proton is located at a distance of about $10^{6}\,\mathrm{cm}$ from the O-point. So the motion of the proton in the vicinity of an O-point in our simulation is indeed non-adiabatic. For the electron, the motion is non-adiabatic when its distance from the O-point is about $10^{4}\,\mathrm{cm}$, close to our numerical grid size. So if we place electrons rather than protons on our MHD background, the motion of the electrons should be fully adiabatic, which could be an explanation why the motion of the electrons is adiabatic in the vicinity of an O-point in Ref.~\onlinecite{Drake2006}. Both adiabatic and non-adiabatic motions of electrons were reported in Ref.~\onlinecite{Fu2006PhPl}, which depend on the scales of gyroradius.

\subsection{Two-dimensional setup versus three-dimensional setup}
 
In this study, we place test particles on a two-dimensional MHD background and investigate the motion of these particles, which is in this case a simplified representation of more realistic situations. In real astrophysical plasmas, there exist no such idealized two-dimensional MHD configurations. Usually, astrophysical plasmas are three-dimensional systems. 
The three-dimensional nature of these plasmas is critically important for understanding magnetic reconnection and particle acceleration~\cite{Khabarova2021SSRv,Pezzi2021SSRv}.
The third-component of $B$ (i.e. the "guide field") plays important roles both on magnetic reconnection and particle motion.
 The three-dimensional reconnection is different from the two-dimensional reconnection in various aspects. It is reported that the three-dimensional reconnection induces turbulence that makes magnetic reconnection fast~\cite{Lazarian1999ApJ} and independent of resistivity~\cite{Lazarian2020PhPl}. The importance of the guide field and fully three-dimensional effects on particle acceleration have been studied by, e.g., Refs.~\onlinecite{Kowal2011ApJ,Li2017A&A,Zhong2016ApJS}. It is also reported in Ref.~\onlinecite{Fu2006PhPl} that electrons can be accelerated in both the X-type and O-type regions without the initial guide field in two-dimensional Particle-In-Cell (PIC) simulations while electrons can only be accelerated in the X-type region when the initial guide field is added. This could be reinvestigated by test-particle simulation in the future.

\subsection{Test-particle approach versus kinetic approach}
 
In our test-particle simulations, the particle motion is fully determined by the MHD background and there are neither interactions between particles nor feedbacks from particles to fluid. The fast-moving charged particles create a current, which generates an electric field counteracting the particle motion and decelerating them. This effect originated from the interactions between particles is not included in the test-particle approach. In a realistic plasma, there are deceleration mechanisms like collision and radiation reaction. The acceleration efficiency should be lowered once these deceleration mechanisms are included. We should add two forces to the right-hand side of Equation (\ref{motion}), i.e., the collisional drag force representing the collisional effects and the Abraham-Lorentz force representing the radiation damping, which should be tested in the future. However, we clearly see how the MHD background influences the particle motion in the test-particle simulation. Especially, we can construct an analytical method to understand the particle energization process in this simplified setup.

MHD is usually an acceptable model to solar and astrophysical plasmas on macroscopic scales. The MHD description is not accurate anymore when scales are comparable with kinetic scales. We now verify the validation of MHD by comparing the scales of our interests with the kinetic scales. 
We first of all calculate some fundamental plasma scales. We are interested in the vicinity of an O-point, so we take the length scale as $l\simeq 10^{6}\,\mathrm{cm}$, the temperature as $T\simeq 10^{7}\,\mathrm{K}$, the magnetic field strength as $B\simeq 1\,\mathrm{G}$, and the number density as $n\simeq 10^{9}\,\mathrm{cm^{-3}}$. Then we obtain the electron Debye length 
\begin{equation}
	\lambda_{D}=(\frac{kT}{4\pi n e^{2}})^{1/2}\simeq 2(\frac{T}{10^{6}\mathrm{K}})^{1/2}(\frac{n}{10^{9}\mathrm{cm^{-3}}})^{-1/2}\,\mathrm{cm}\simeq 6\,\mathrm{cm},
\end{equation}
the electron gyroradius 
\begin{equation}
r_{g\mathrm{e}}=\frac{c}{eB}(m_{\mathrm{e}}kT)^{1/2}\approx 2(\frac{B}{10\,\mathrm{G}})^{-1}(\frac{T}{10^{6}\,\mathrm{K}})^{1/2}\simeq 60\,\mathrm{cm},	
\end{equation}
the proton gyroradius 
\begin{equation}
	r_{g\mathrm{p}}=\frac{c}{eB}(m_{\mathrm{p}}kT)^{1/2}\simeq 10^{2}(\frac{B}{10\,\mathrm{G}})^{-1}(\frac{T}{10^{6}\,\mathrm{K}})^{1/2}\simeq 3000\,\mathrm{cm},
\end{equation}
the electron inertial length 
\begin{equation}
r_{i\mathrm{e}}=c(\frac{m_{\mathrm{e}}}{4\pi n e^{2}})^{1/2}\approx 30(\frac{n}{10^{9}\mathrm{cm^{-3}}})^{-1/2}\simeq 30\,\mathrm{cm},	
\end{equation}
and the proton inertial length 
\begin{equation}
r_{i\mathrm{p}}=c(\frac{m_{\mathrm{i}}}{4\pi n e^{2}})^{1/2}\approx 10^{3}(\frac{n}{10^{9}\mathrm{cm^{-3}}})^{-1/2}\simeq 1000\,\mathrm{cm}.
\end{equation}
 It should be noted that the above electron gyroradius and proton gyroradius are calculated based on the thermal velocity of the electrons and ions, which are different from the gyroradius of our test-particles. The scale of our interest $l\simeq 10^{6}\,\mathrm{cm}$ is much larger than the above plasma typical scales. In this sense, our MHD simulation is still acceptable at a scale of $l\simeq 10^{6}\,\mathrm{cm}$, at which the test-particles are accelerated. 
 However, when the particles are energized, the gyroradius of the particles are larger than $l\simeq 10^{6}\,\mathrm{cm}$. Thus the test-particles can not be treated as a fluid. Our MHD simulation only describes the behaviors of thermal particles that can be treated as a fluid. The high energy non-thermal particles with a large gyroradius are not described by MHD. In this sense, our MHD model is neither a self-consistent nor a complete description of the plasma system. However, our model shows the possibility of acceleration in the vicinity of an O-point in sub-gyroradius scale, which only relies on magnetic configuration and can be tested by kinetic studies.

\section{Conclusions}\label{conclusion}
In this study, we investigate magnetic reconnection during island merging and particle acceleration inside plasmoids. The research can be summarized as follows:

\begin{enumerate}
\item
The minimum Lundquist number required to trigger plasmoid instability is $2.9\times10^{4}$, which is consistent with the results in Ref.~\onlinecite{Huang2010PhPl}. 
\item
The motion of a proton in our test-particle simulations on a time-fixed background can be divided into 3 phases, a guiding center drift phase, a phase of adiabatic motion, and a phase of relativistic non-adiabatic motion around the vicinity of an O-point. The proton energy is almost conserved during guiding-center drift and adiabatic motion while protons can be accelerated to hundreds of GeVs within 30 seconds during the non-adiabatic motion in the vicinity of an O-point.
\item
Protons can gain energy from the electric field during the non-adiabatic motion in the vicinity of an O-point. In our simplified two-dimensional setup, the electric field consists of a convective part $E_{con}=(1/c)(u_{y}B_{x}-u_{x}B_{y})$ and a resistive part $E_{res}=\eta J_{z}$. In the high-magnetic-Reynolds-number case, the convective part $E_{con}$ dominates. Due to deviation of generic magnetic islands from perfect ellipses and the non-uniformity of the fluid velocity field $\mathbf{u}$, the electric field around an O-point is non-symmetrically distributed. Protons move back and forth around the O-point. The symmetric electric field does zero work to the proton during each cycle of this motion. However, the proton gains a small amount of energy during each cycle of motion from the non-symmetric electric field. The long-term ($\sim 30\,\mathrm{s}$) accumulation of the energy gained during each cycle of motion can lead to relativistic energies for these accelerated particles.  
\item
The energy gained during each cycle of bouncing motion of the proton is proportional to the ratio of fluid velocity and light speed, i.e. $v\sim O(u/c)$, which is similar to the first-order Fermi-type acceleration in the sense that the gain is proportional to the first power of $u/c$.  
\item
Protons with speeds close to $\sqrt{2}c/2\approx0.7c$ are accelerated with the highest efficiency. 
\item
The trajectories of the bouncing motion of a proton in the {\it x-y} plane are close to Lissajous curves.
\item
The long-term motion of a proton in the {\it z}-direction is equivalent to the motion of a charged particle in a constant uniform electric field directed along the {\it z}-axis.
\item
The role of O-points during particle acceleration is two-fold, either scattering particle away or accelerating particles.
\end{enumerate}

\begin{acknowledgments}
We acknowledge support by a joint FWO-NSFC grant G0E9619N. RK received funding from the European Research Council (ERC) under the European Unions Horizon 2020 research and innovation programme (grant agreement No. 833251 PROMINENT ERC-ADG 2018), and from Internal Funds KU Leuven, project C14/19/089 TRACESpace.
The computational resources and services used in this work were provided by the VSC (Flemish Supercomputer Center), funded by the Research Foundation--Flanders (FWO) and the Flemish Government--department EWI.
FB is partially supported by a Junior PostDoctoral Fellowship (grant number 12ZW220N) from Research Foundation--Flanders (FWO).
\end{acknowledgments}
 
\medskip
\textbf{\large Author Declarations}

\textbf{\small Conflicts of Interest}
The authors have no conflicts to disclose.

\medskip
\textbf{\large Data availability}

The data that support the findings of this study are available from the corresponding author upon reasonable request.

%

 
\end{document}